\documentclass[12pt,a4paper]{article}
\pdfoutput=1
\usepackage{jheppub}
\usepackage{amsmath,amssymb,amsbsy,amsfonts,latexsym,graphicx,hyperref}
\usepackage{color,array,subfigure} 
\usepackage{color,eucal} 
\usepackage[default]{dhucs-interword}
\usepackage[hangul]{dhucs-setspace}
\usepackage{multirow, graphicx,url,mathrsfs}
\usepackage{wrapfig,boxedminipage,setspace,subfigure,epsfig}
\usepackage{amsxtra,amstext,latexsym,dsfont}
\usepackage{amsmath,amssymb,amsbsy,amsfonts,latexsym,hyperref}
\usepackage{color,array,subfigure}
\usepackage{graphicx,amsmath,amssymb}
\usepackage{cancel}
\usepackage{txfonts}

\newcommand{\be}{\begin{equation}}
\newcommand{\ee}{\end{equation}}
\newcommand{\bea}{\begin{eqnarray}}
\newcommand{\eea}{\end{eqnarray}}
\newcommand{\beq}{\begin{equation}}
\newcommand{\eeq}{\end{equation}}
\newcommand{\beqa}{\begin{eqnarray}}
\newcommand{\eeqa}{\end{eqnarray}}
\newcommand{\curl}{\nabla\times}

\newcommand{\ket}[1]{\mbox{$| #1 \rangle$}}

\title{\Large Stability of Topology in interacting Weyl Semi-Metal, \\  and  Topological Dipole in Holography}
\author[a]{Geunho Song}
\author[a]{Junchen Rong} 
\author[a]{and Sang-Jin Sin}
\emailAdd{sgh8774@gmail.com}
\emailAdd{junchenrong@gmail.com}
\emailAdd{sangjin.sin@gmail.com}
\affiliation[a]{ Department of Physics, Hanyang University, Seoul 133-791, Korea }
\abstract{We discuss the stability of the topological invariant of the strongly interacting Weyl semi-metal at finite temperature. 
 We find that if the   interactions and temperature of the   system are controlled by the  holography, the topology  is stable even in the case the Fermi surface become fuzzy. 
We  give an argument to show that although the self energy   changes  the  spectral  function significantly to make the Fermi surface fuzzy, it   cannot change the singularity structure of the Berry phase, which leads to the stability of the topology.   
 We also find that  depending on the mass term structure of the fermion Lagrangian,  topological dipoles can be created.  }
\subheader{\today
  }

\arxivnumber{}

\begin{document}

\maketitle

\section{Introduction} 
Topological matter\cite{Kane2005, Qi2008, Qi2009, Raghu2008, Witten2015}  is a new quantum state of matter that has a promising application for quantum computations\cite{freedman2003topological,nayak2008non} and there has been a flurry of activities in   last 10 years.  It is topological since the Hilbert space has a non-trivial topological structure and the key is a non-trivial edge state associated with it. 
It started with materials with negligible interaction, but recently  the importance of its existence in the presence of strong interaction and finite temperature is  getting much attention\cite{ Raghu2008, Fidkowski2010,Gurarie2011, Wang2012, Landsteiner2016a, Landsteiner2016}. 
The basic question is whether the topological structure, which has been discovered in the non-interacting case, can survive when one turns on the interaction or other deformations of the system like temperature or pressure. 
One can also ask whether a new topological structure which was absent in the weakly interacting case can arise due to the strong interaction. 
The purpose of this paper is to answer both of the question affirmatively. We will show that  the topological structure for Weyl semi-metal is robust even in the case when the spectral function shows that the line width is broaden and band structure is fuzzy. 
Also we will describe a model with topological dipoles, where Weyl points are separated by only a small distance in momentum space. 
We will see that such objects are not so stable in the sense that they can disappear as temperature goes up high enough. 

The general definition of the topological invariant for interacting system  is already defined in terms of the full Green's function in \cite{Wang2012,Qi2008, Gurarie2011}. However, it can not be very useful  unless one can actually calculate the full Green function, which is beyond the perturbative field theory. 
Here we utilize the holographic setup to  calculate the Green functions and use the result to construct the effective Hamiltonian, which in turn allows us to calculate the winding number of the Weyl points.  
Previously the Weyl semi-metal in holographic set up was discussed in \cite{Y.liu2018}, and topological invariant was proved to be well define in the limit of zero temperature and small fermion mass. 
Here we extend it to the finite temperature and finite mass, where spectral function becomes   fuzzy due to the large imaginary part of self energy which gives the line broadening. 

In section 2, we will set up the problem by reviewing the Weyl semi-metal in quantum field theory and  the holographic version of Weyl semi-metal. We will also give spectral function at finite temperature. 
In section 3, we examine the stability of topological invariant. 
In section 4. We define and study a model for topological dipole. 

\section{Weyl semi-metal in QFT and holography}
\subsection{Weyl semi-metal in quantum field theory of 3+1 dimension}
\begin{figure}[ht!]
\centering
    \subfigure[$({B_z},M)=(3,0)$ ]
    {\includegraphics[width=35mm]{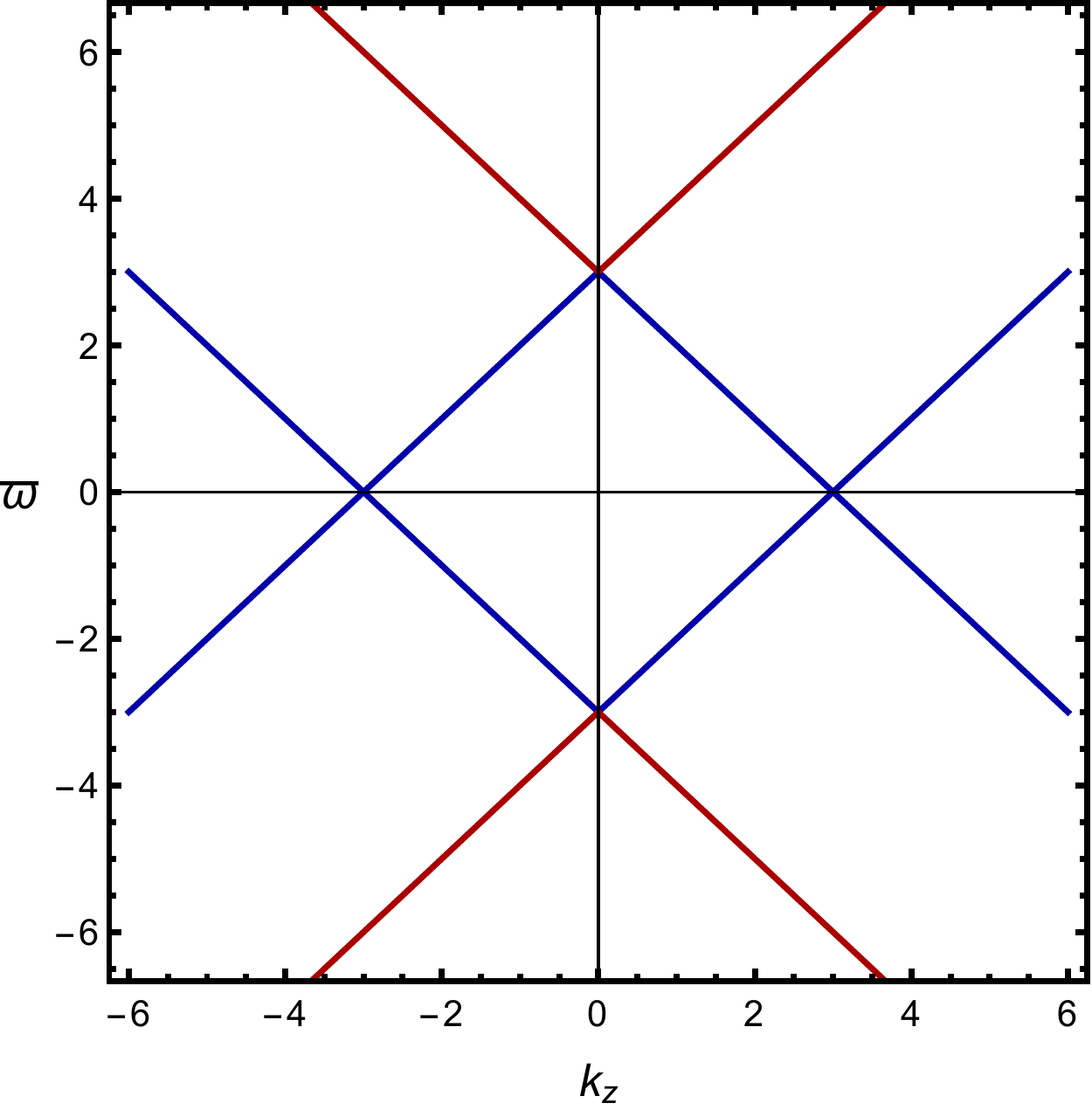}  }
   \subfigure[$({B_z},M)=(3,1)$ ]
    {\includegraphics[width=35mm]{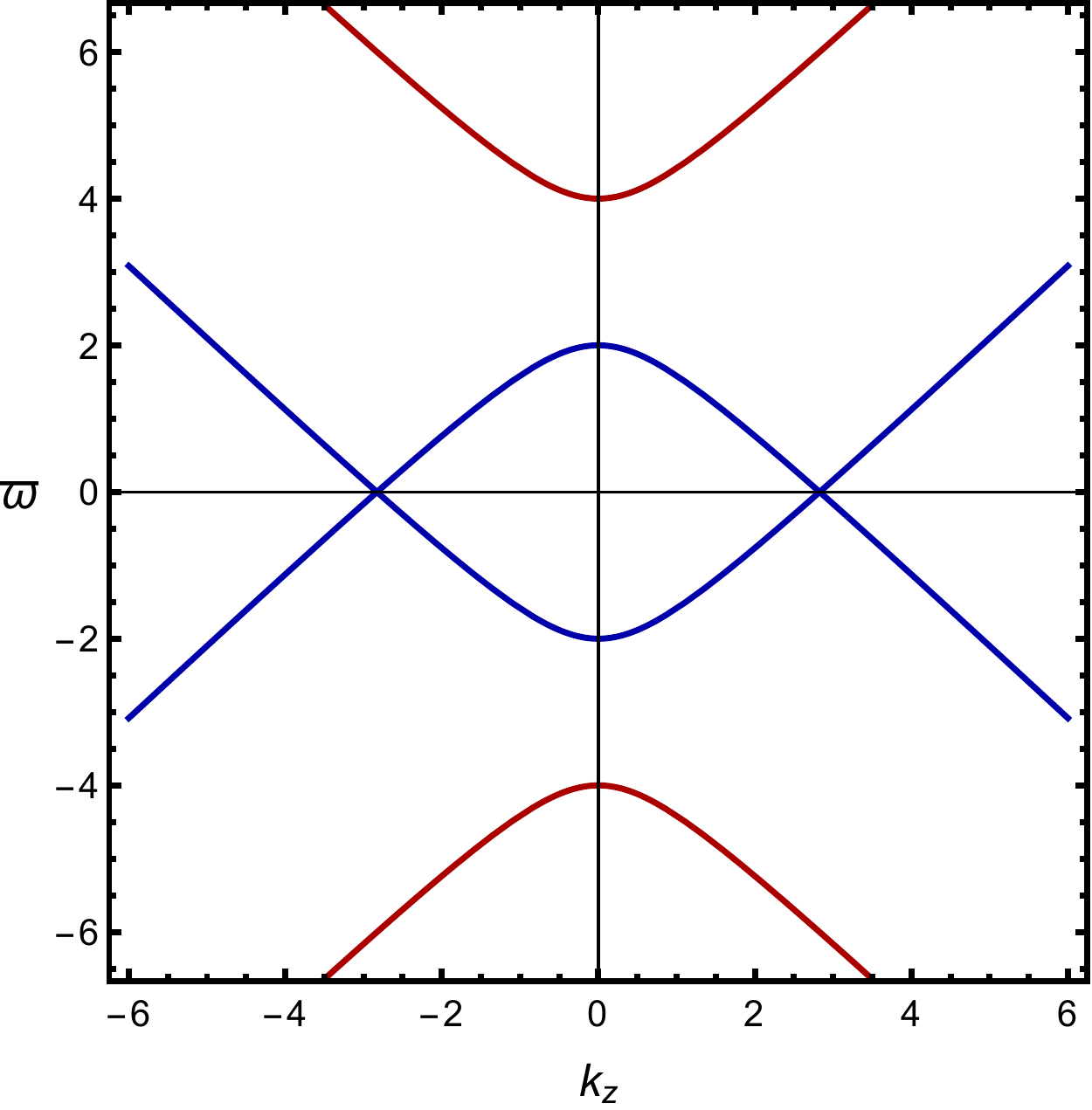}  }
     \subfigure[$({B_z},M)=(3,3)$ ]
    {\includegraphics[width=35mm]{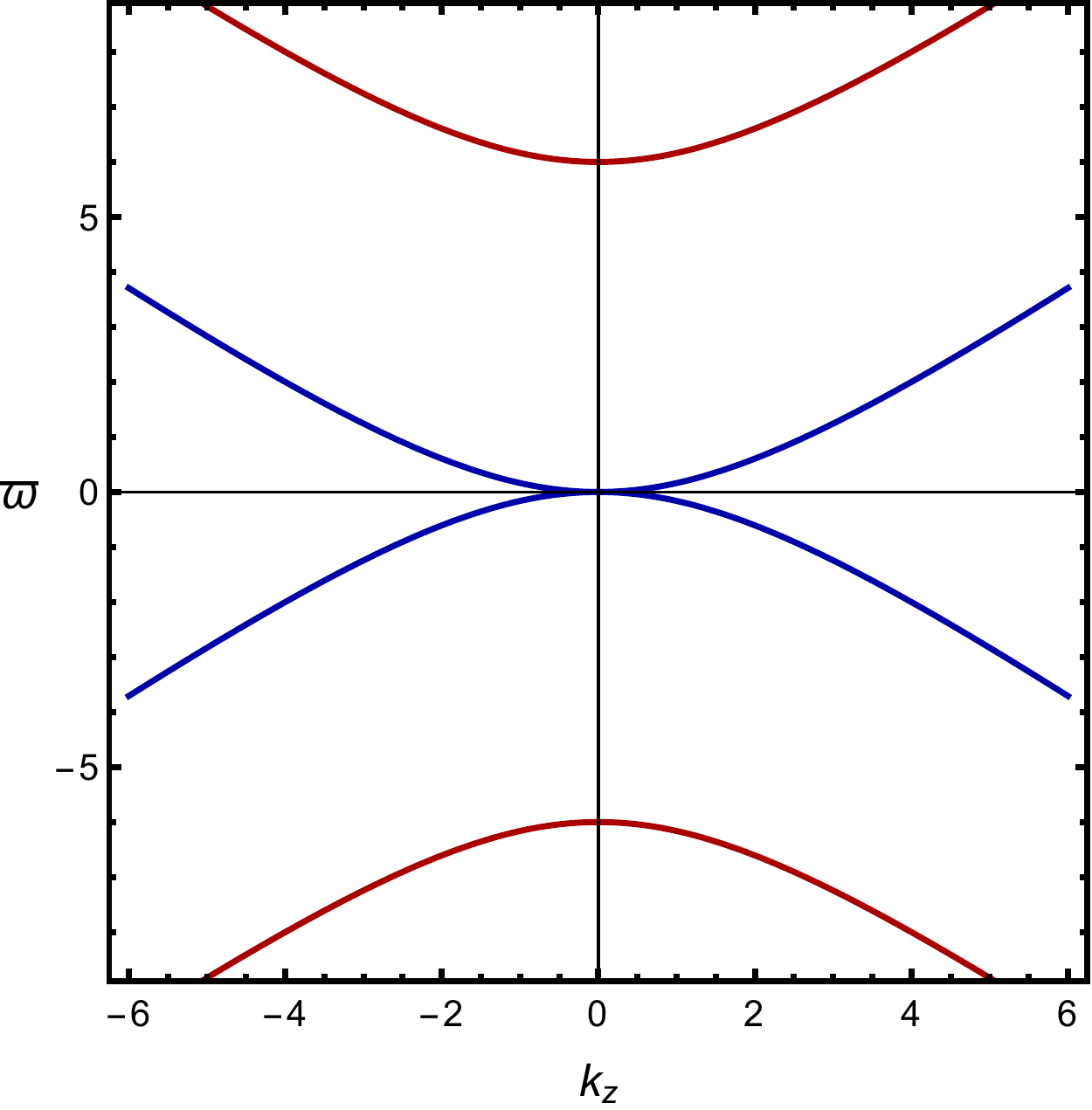}  }
     \subfigure[$({B_z},M)=(3,5)$ ]
    {\includegraphics[width=35mm]{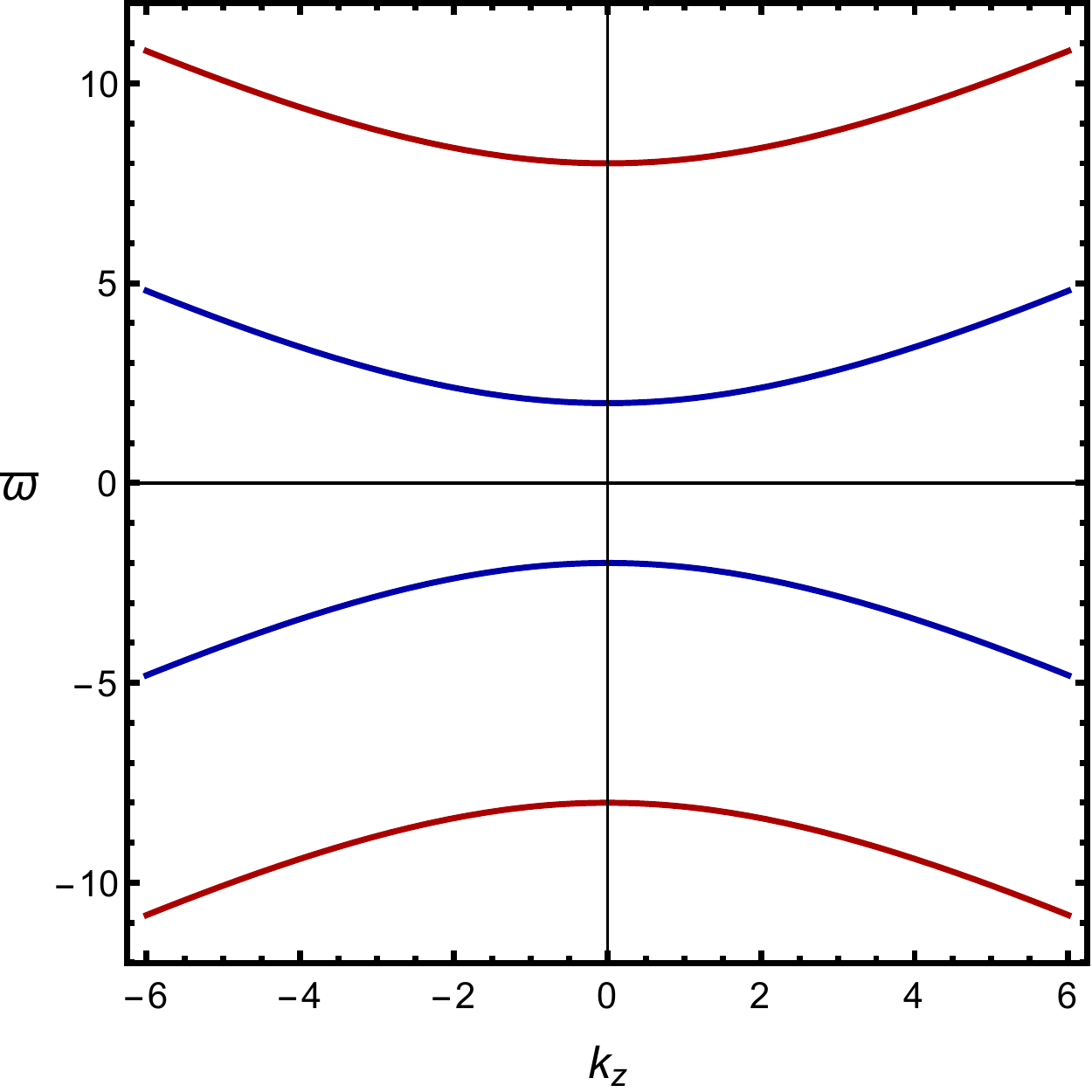}  }
              \caption{(a)-(d):The band structure depends on $B_{z}/M$.  The figure is in $(k_z,\bar{\omega})$-space at  $k_x=0,k_y=0$ slice. Here $\bar{\omega}=\omega+q\mu$. 
         } \label{dspaxialz}
\end{figure}

Here we briefly review a quantum field theoretical (QFT)  model for Weyl semi-metal (WSM). WSM has the separate band crossing points in momentum space which can be achieved by breaking time-reversal symmetry of  Dirac semi-metal. 
Consider   the fermion action in (3+1) dimensional Minkowski space-time with axial vector interaction  ~\cite{Goswami2013,Colladay1998}: 
\begin{align}
	S=\int d^4x \bar{\Psi} i(\cancel{\partial}-q\cancel{A}-M-iB_{\mu}\gamma^5\gamma^{\mu})\Psi
\end{align} 
where $A=\mu dt$ with the chemial potential $\mu$. Expanding $\Psi$ in momentum basis $e^{-i(\omega t-\mathbf{k}\cdot\mathbf{x})}\psi$, the equation of motion is given by
\begin{align}
	\left(i \cancel{K}-M -i  {B_{\nu}}\gamma^5\gamma^{\nu}\right)\psi=0\label{axialspace}
\end{align}
where $K=(\omega+q \mu,\vec{k})$, and index $\nu$ is not summed in (\ref{axialspace}), that is, $B_{\nu}$ is just coefficient of $\gamma^{\nu}$.
For   simplicity, we choose the configuration with only  $B_{z}$  non-zero. Then the dispersion relations has four branches given by
\begin{align}
		 {B_z}\  :\  \bar{\omega}=\omega+q\mu=\pm\sqrt{ {B_z}^2+\vec{k}^2+M^2 \mp2\sqrt{{B_z}^2(k_z^2+M^2)}},
		\label{dspaxspace}
%		  &\quad\pm\sqrt{{B_z}^2+\vec{k}^2+m^2+2\sqrt{{B_z}^2(k_z^2+m^2)}}\label{dspaxspace2}
\end{align}
%For (\ref{dspaxspace}), we can obtain only 
For $|{B_z}|>M$, the band crossing happens at $(k_x, k_y, k_z)=(0,0, \pm\sqrt{{B_z}^2-M^2})$ and the spectrum is gapless. The seperation  between the crossing points is   $2\bar{B}_{eff}=2\sqrt{ {B}_z^2-M^2}$. See Figure \ref{dspaxialz}. 

\begin{figure}[ht!]
\centering
    \subfigure[$\bar{\omega}=1$ ]
    {\includegraphics[width=45mm]{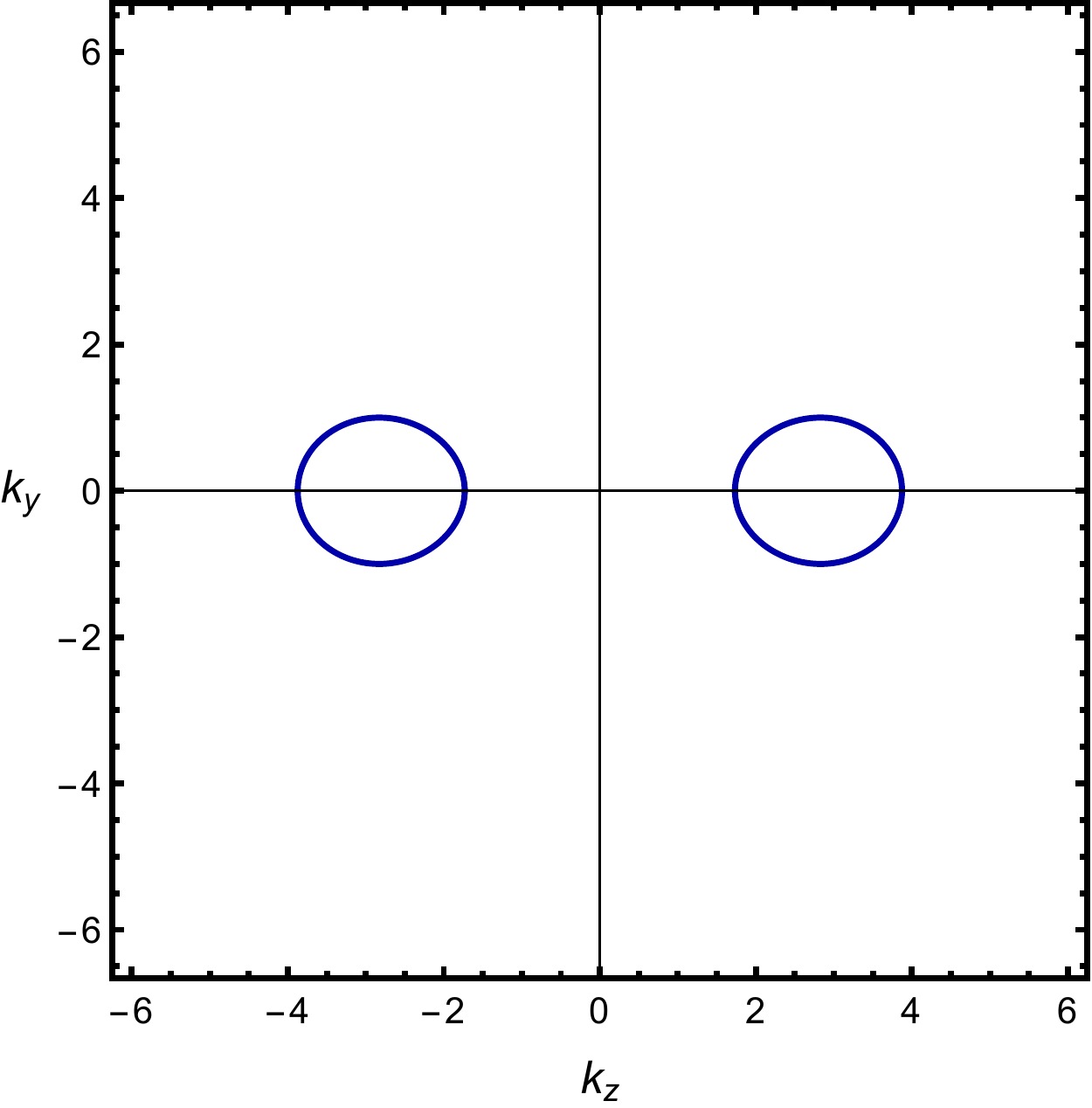}  }
   \subfigure[$\bar{\omega}=2$ ]
    {\includegraphics[width=45mm]{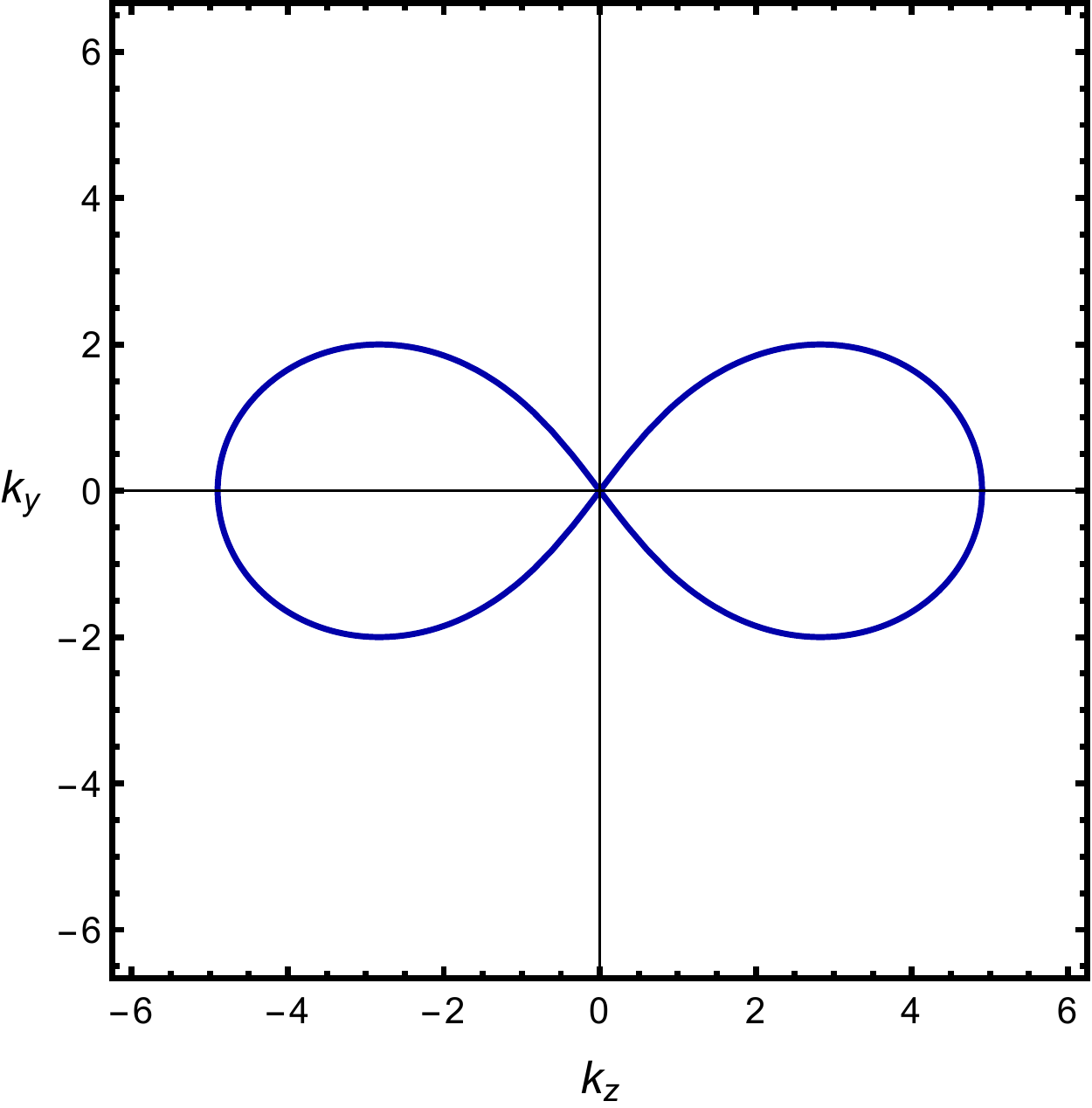}  }
     \subfigure[$\bar{\omega}=3$ ]
    {\includegraphics[width=45mm]{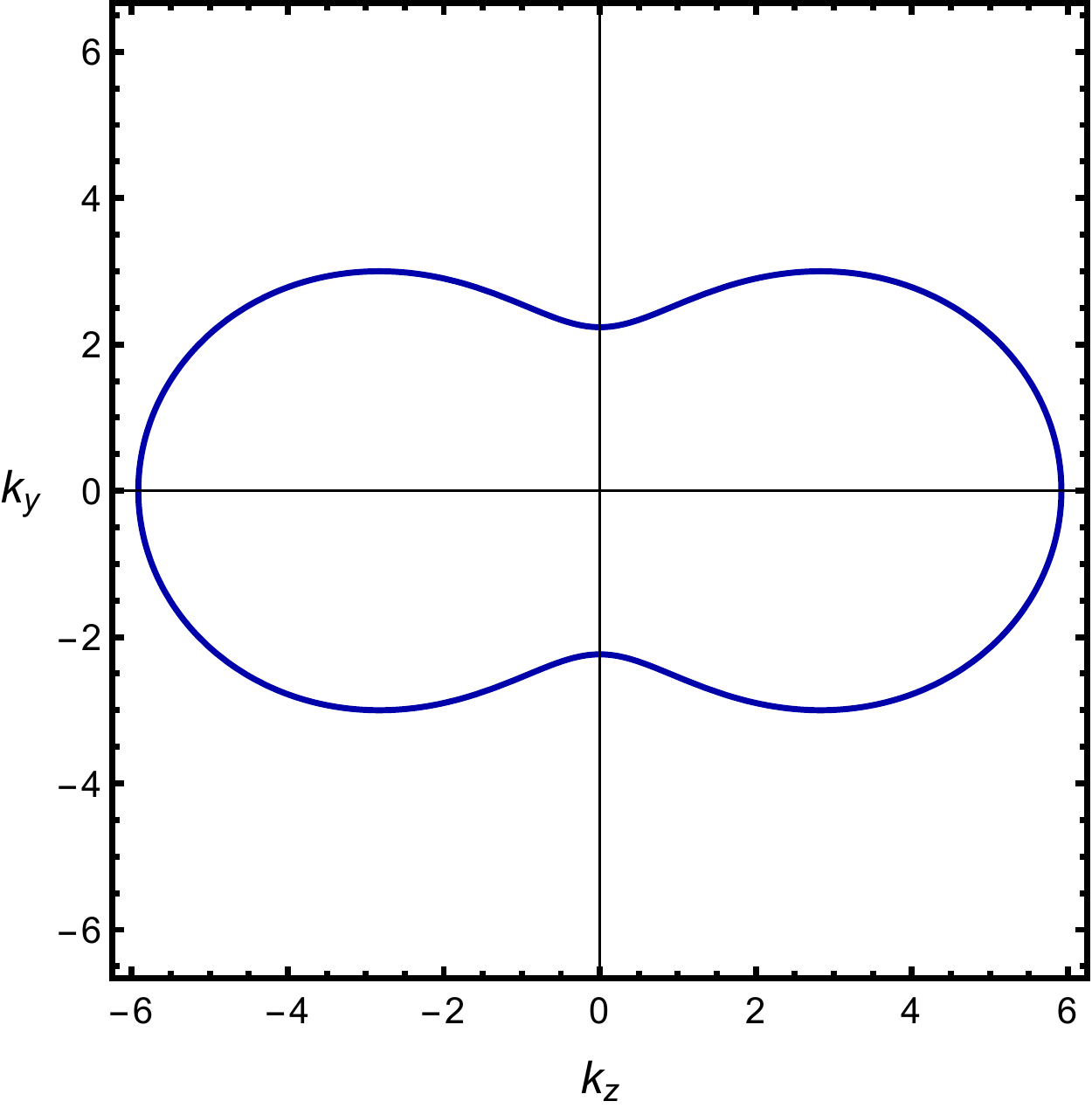}  }
              \caption{(a)-(c): The different  $\bar{\omega}$  slices of the band structure  at  $k_x$=0  with $(\bar{B}_z,M)=(3,1)$. The radius of circle decreases as $\bar{\omega}$ apparoches to 0. This result is the same for $k_y=0$ plane.          } \label{dspaxialzzy}
\end{figure}
 
 On the other hand,  for $|{B_z}|<M$, a gap  opens   and  its size is given by $2\Delta=2(M- {B}_z)$. Figure \ref{dspaxialzzy} shows the top-view of the fermion spectrum with  sections of Dirac cones  which has the centers at the band crossing point of spectrum and it shrinks as we approach to $\bar{\omega}=0$, which implies the spectrum forms cone-structure near the band crossing points.
 
\subsection{Holographic Fermions and their spectral function}
\begin{figure}[ht!]
\centering
    \subfigure[Profile for $A_z$ ]
    {\includegraphics[width=60mm]{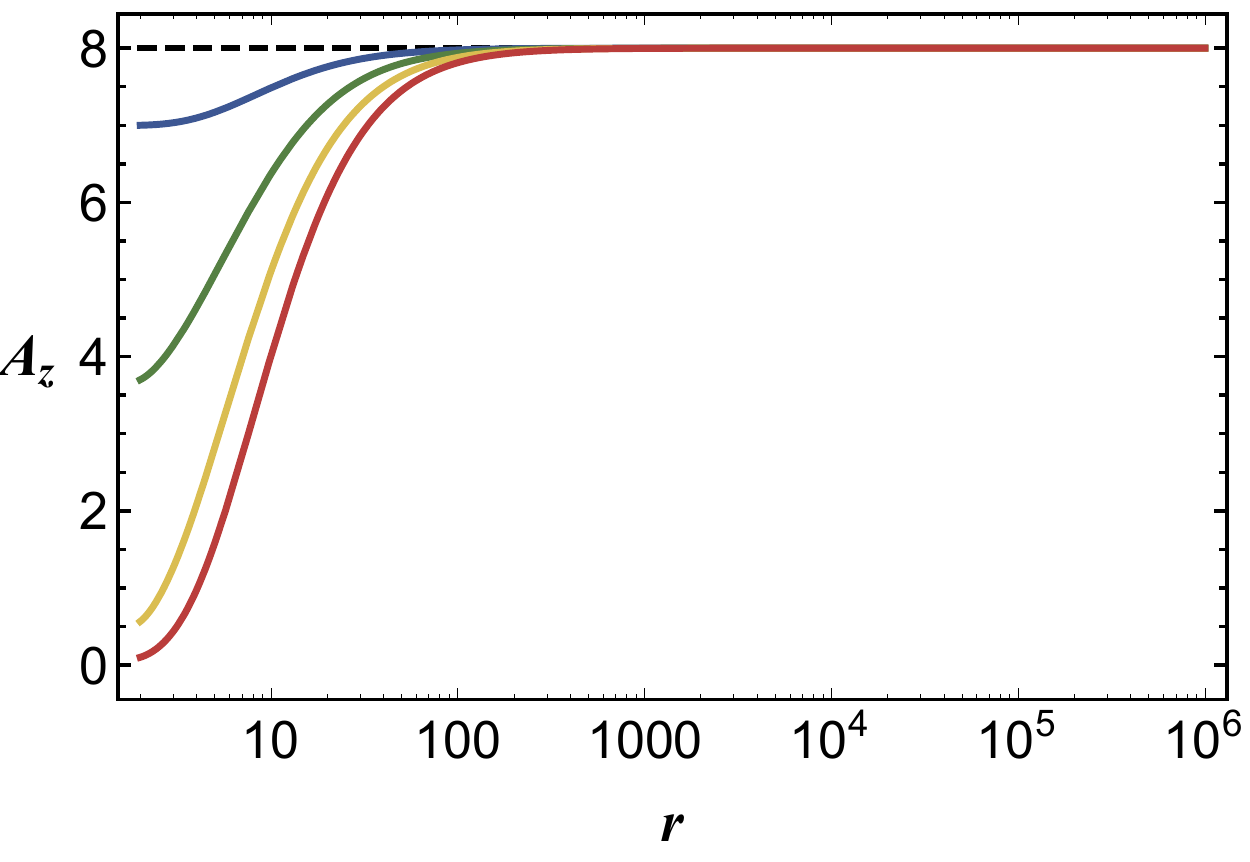}  }
   \subfigure[Profile for $\Phi$ ]
    {\includegraphics[width=60mm]{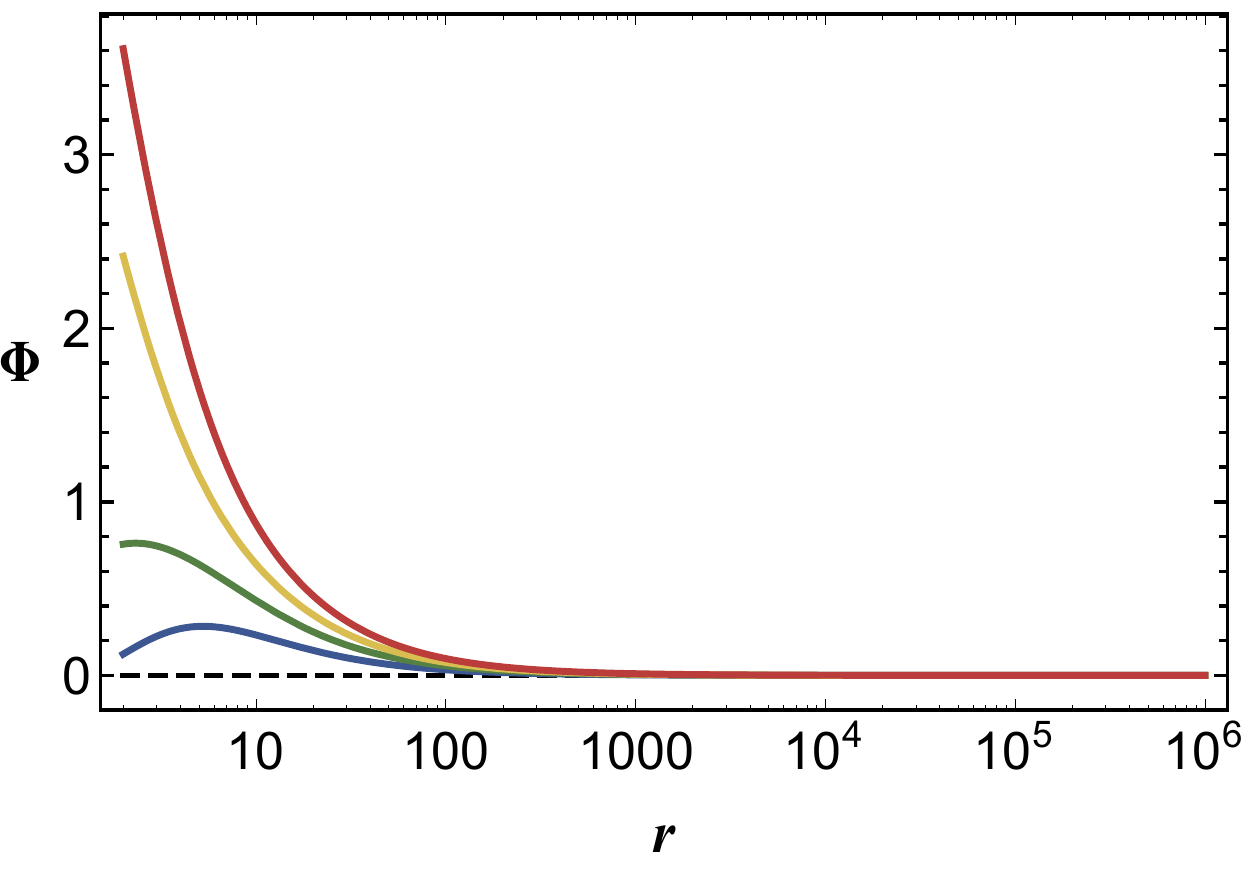}  }  
              \caption{(a)(b) Profiles for $A_z$ and $\Phi$ where we fix $b=8$ and $T=2/\pi$. We use  $M=0$, $3.3,$ $5.5$, $7.5$, $9.7$ for flat-dashed, blue, green, yellow, red (from top to bottom in (a) and  from bottom to top in (b)).
         } \label{fig:profile}
\end{figure}

To reproduce the above  band structure of  Weyl-semi metal in the holographic set up, we use a model which  was first introduced  in~\cite{Y.liu2018} 
\begin{align} \label{Action}
 S&=S_1+S_2+S_{int}\\
 S_1&=\int d^5x\sqrt{-g}i\bar{\Psi}_1(\Gamma^aD_a-m_f-iA_a\Gamma^a)\Psi_1,\nonumber\\
 S_2&=\int d^5x\sqrt{-g}i\bar{\Psi}_2(\Gamma^aD_a+m_f+iA_a\Gamma^a)\Psi_2,\nonumber\\
 S_{\text{int}}&=\int d^5x\sqrt{-g}(i\eta_1\Phi\bar{\Psi}_1\Psi_2+i\eta_1^*\Phi^*\bar{\Psi}_2\Psi_1),\nonumber
 \end{align}
where $\mathcal{D}_M=\partial_M+\frac{1}{4}\omega_{\underline{a}\underline{b}M}\Gamma^{\underline{a}\underline{b}}$ 
is the  the covariant derivative and
  $\omega_{\underline{a}\underline{b}M}$ is the bulk spin connection,    $\Gamma^{\underline{a}\underline{b}}=\frac{1}{2}\lbrack \Gamma^{\underline{a}},\Gamma^{\underline{b}} \rbrack$. 
  $A_a$ is a gauge field with zero bulk mass and $\Phi$  is a scalar with $m^{2}_{\Phi}=-3$, breaking the time reversal symmetry(TRS) and chiral symmetry respectively.  $M$ denotes bulk spacetime indices and $a,b$ denote bulk tangent space ones. 
  In ref. ~\cite{Y.liu2018}, zero temperature analysis was done. Here we will consider the finite temperature case.  For this purpose, 
 we take the Schwarzschild-$AdS_5$ background (\ref{AdS5}).  
 \begin{align}
ds^2 &= -r^2f(r) dt^2 +\frac{1}{r^2f(r)} dr^2 +\frac{r^2}{L^2}d\vec{x}^2  \cr
f(r) &= \frac{1}{L^2}\left(1-\frac{r_0^4}{r^4}\right) \label{AdS5}
\end{align} 
where $L$ is $AdS_5$ radius and $r_0$ is the radius of the black hole which defines the temperature of boundary theory, where $T=f'(r_0)/4\pi=r_0/\pi L^2$. 
%In ~\cite{Y.liu2018}, they consider the case at zero temperature only, while we are dealing with the finite temperature case,   there's not much difference.
For $A_z$ and $\Phi$, we have the equations of motion\cite{Landsteiner2016} as follows:
\begin{align}\label{eqphia}
	A_z''+\left(\frac{3}{r}+\frac{f'}{f}\right)A_z'-\frac{2\Phi^2}{r^2f}A_z&=0\\
	\Phi''+\left(\frac{5}{r}+\frac{f'}{f}\right)\Phi'-\left(\frac{A_z^2}{r^4f}+\frac{m_{\Phi}^2}{r^2f}\right)&=0
\end{align}
We can introduce the parameter $b$ and $M$ as a boundary condition for the fields $A_z$ and $\Phi$ which satisfy (\ref{eqphia})
\begin{align}
	\lim_{r\rightarrow\infty}A_z(r)=b,\qquad  \lim_{r\rightarrow\infty}r\Phi(r)=M
\end{align}
 The specified parameter $(b,M)$ at the boundary  can be reached by choosing proper horizon values of $A_{z}$ and $\Phi$, whose profles are shown in Figure \ref{fig:profile} (a) and (b) respectively. 

\begin{figure}[ht!]
\centering
      \subfigure[$(b,M)=(8,0)$ ]
    {\includegraphics[width=35mm]{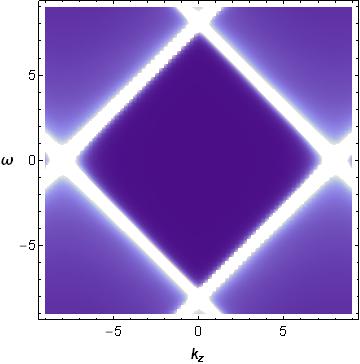}  }
      \subfigure[$(b,M)=(8,3.3)$ ]
    {\includegraphics[width=35mm]{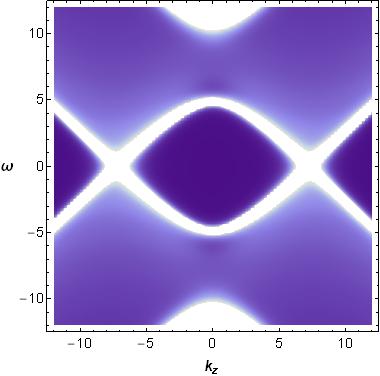}  }
     \subfigure[$(b,M)=(8,7.5)$ ]
    {\includegraphics[width=35mm]{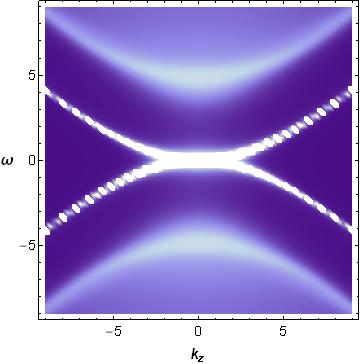}  }
      \subfigure[$(b,M)=(8,9.7)$ ]
    {\includegraphics[width=35mm]{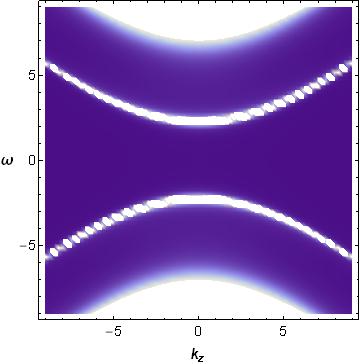}  }
              \caption{(a)-(d):Spectral densities on $(k_z,\omega)$-space with $k_x=k_y=0$ at $ T=2/\pi$.  Separation between the Weyl points is approximately $2\sqrt{b^{2}-M^{2}}$. 
         } \label{spectralliu}
\end{figure}

\begin{figure}[ht!]
\centering
    \subfigure[$T=2/\pi$ ]
    {\includegraphics[width=45mm]{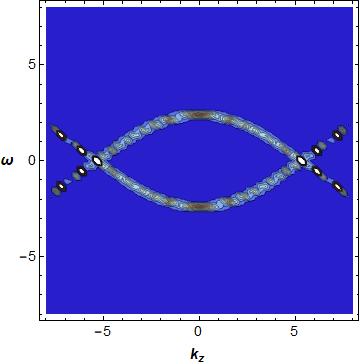}  }
   \subfigure[$T=10/\pi$ ]
    {\includegraphics[width=45mm]{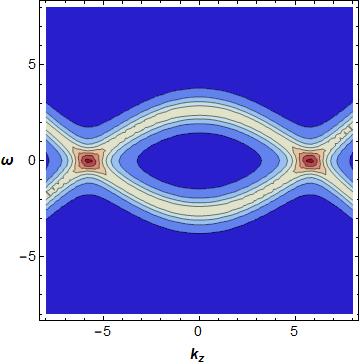}  }
     \subfigure[$T=20/\pi$ ]
    {\includegraphics[width=45mm]{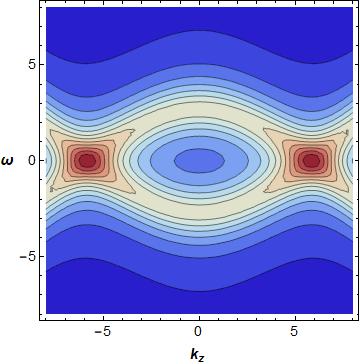}  }
              \caption{(a)-(c):  
              As temperature increases, spectrum broadens.  Color code denotes the spectral height.   
              %$(k_z,\omega)$-space with
              We used $k_x=k_y=0$ and $(b,M)=(8,3.3)$. 
    } \label{tempevoliu}
\end{figure}

\begin{figure}[ht!]
\centering
% \subfigure[Phase diagram]
    {\includegraphics[width=60mm]{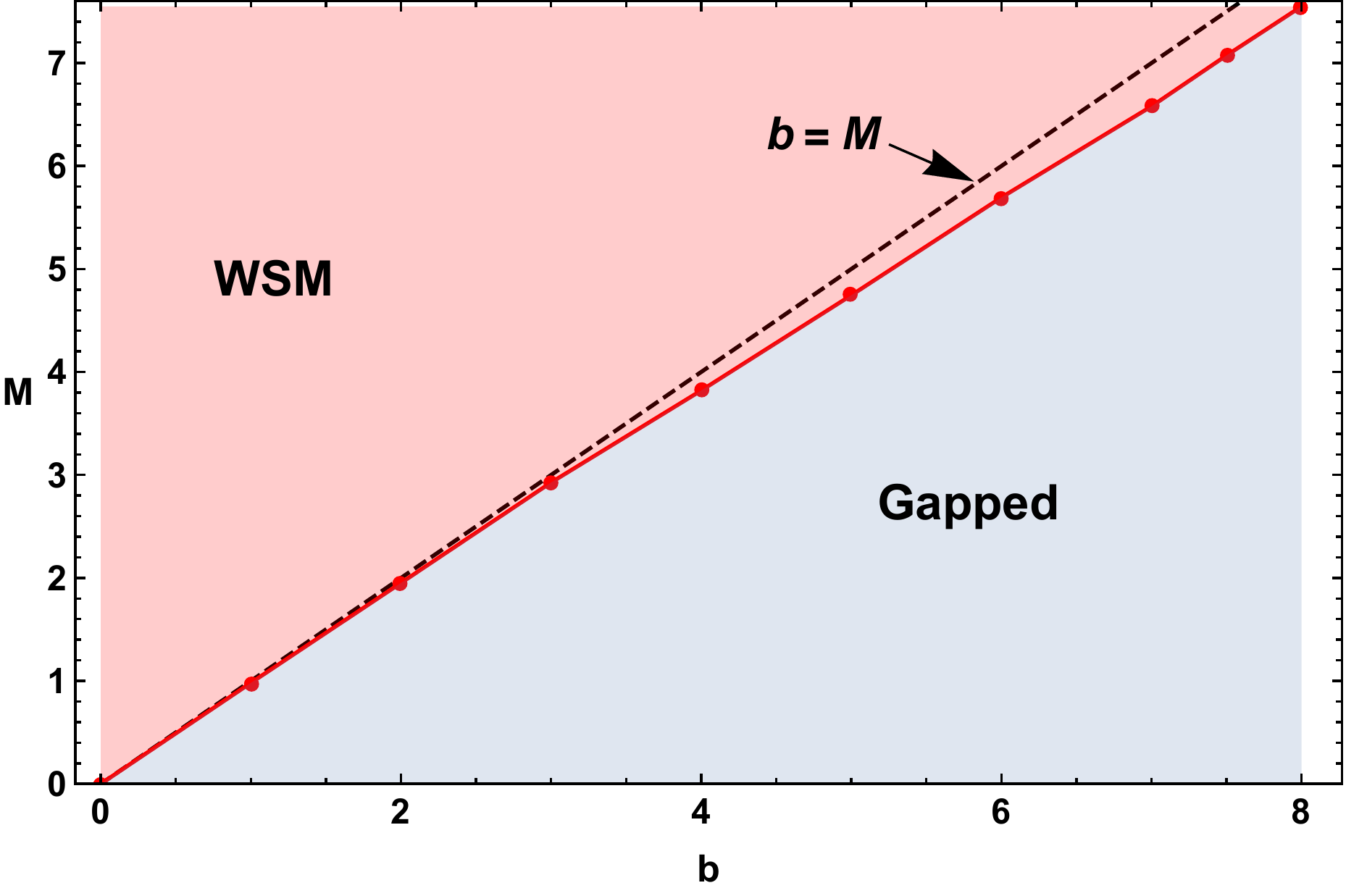}  }
              \caption{ Phase diagram in $(b,M)$ space. We used $T=2/\pi$. Unlike the field theory,  $M_c/b=1$ holds only approximately. Dashed line is  for  $b=M$.
         } \label{phaseD}
\end{figure}

We use the convention of $\Gamma$-matrices for fermion action as follows:
\begin{align}
	 \Gamma^{\underline{t}}=\left(\begin{array}{cc}0&\gamma^{t}\\   \gamma^{t}&0\end{array}\right),\quad
	 \Gamma^{\underline{i}}=\left(\begin{array}{cc}0&\gamma^{i}\\   -\gamma^{i}&0\end{array}\right),\quad	
	 \Gamma^{\underline{r}}\equiv\Gamma^5=\left(\begin{array}{cc}1&0\\  0&-1\end{array}\right),\quad
	  (\gamma^{t},\gamma^i)=i(\mathbf{1}_2,	\sigma^i)		 
\end{align}
where $ \Gamma^M=e^M_{\underline{a}}\Gamma^{\underline{a}}$ and $e^M_{\underline{a}}$ is the inverse vielbein.
%The mass $M$ of the boundary fermion is  encoded in 
%the solution of $\Phi$ by 
 Taking $\eta_1=1$, the equations of motions are given by
 \begin{align}
 	(\Gamma^a D_a-m_f-iA_z\Gamma^z)\Psi_1-\Phi\Psi_2&=0\nonumber\\
 	(\Gamma^a D_a+m_f+iA_z\Gamma^z)\Psi_2-\Phi\Psi_1&=0
 \end{align} 
% Notice that $\Phi=M/r$ is the source term  encoding the boundary mass $M$. We will study the effect of the  condensation operator $\Phi={\tilde M}/r^3$ later.  
Expanding $\bar{\Psi}_l$ in Fourier space,
\begin{align}
	\Psi_l=(-gg^{rr})^{-1/4}e^{ik_{\mu}x^{\mu}}\psi_l,
\end{align}
with $l=1,2$,  the equations of motion for fermions become \begin{align}
\left(\sqrt{g^{rr}}\Gamma^{\underline{r}}\partial_r+\sqrt{g^{tt}}\Gamma^{\underline{t}}(-i\omega)+i\sqrt{g^{ii}}(k_x\Gamma^{\underline{x}}+k_y\Gamma^{\underline{y}}+(k_z\mp A_z)\Gamma^{\underline{z}})+(-1)^lm_f\right)\psi_l-\Phi\psi_{3-l}=0
\end{align}
where we fixed $L=1$.
Near the boundary $r\rightarrow\infty$, the spinors behave as
\begin{align}
\psi_1^{T}&=\left( 
		A_1^1\  r^{m_f}, A_2^1 \ r^{m_f}, A_3^1\ r^{-m_f}, A_4^1\ r^{-m_f} \right)+\cdots,\qquad \\ 
\psi_2^{T} &=\left(A_1^2\ r^{-m_f}, A_2^2 r^{-m_f}, A_3^2\ r^{m_f}, A_4^2\ r^{m_f}		\right)+\cdots.		
\end{align}
We have 8 variables of two first order dirac equations so that 8 ``initial'' conditions are required for radial evolution.  Eliminating outgoing conditions at the horizon, the degrees of freedom are reduced to half.  We choose four different initial conditions   at the horizon and solve the equations to get  near boundary values, which determines the retarded Green functions. We denote each initial conditions as I, II, III, IV respectively. We can construct the source and expectation matrices as follows:
\begin{align}
	\mathbf{A}=\left(\begin{array}{cccc}
		A_1^{1,I} &A_1^{1,II}&A_1^{1,III}&A_1^{1,IV}\\
		A_2^{1,I} &A_2^{1,II}&A_2^{1,III}&A_2^{1,IV}\\
		A_3^{2,I} &A_3^{2,II}&A_3^{2,III}&A_3^{2,IV}\\
		A_4^{2,I} &A_4^{2,II}&A_4^{2,III}&A_4^{2,IV}\\
	\end{array}\right),\qquad 
	\mathbf{D}=\left(\begin{array}{cccc}
		-A_1^{2,I} &-A_1^{2,II}&-A_1^{2,III}&-A_1^{2,IV}\\
		-A_2^{2,I} &-A_2^{2,II}&-A_2^{2,III}&-A_2^{2,IV}\\
		A_3^{1,I} &A_3^{1,II}&A_3^{1,III}&A_3^{1,IV}\\
		A_4^{1,I} &A_4^{1,II}&A_4^{1,III}&A_4^{1,IV}\\
	\end{array}\right)
\end{align}
The Green function can be obtained by $\mathcal{G}_R=i\Gamma^t\mathbf{D}\mathbf{A}^{-1}$. See Appendix for more details. 
The spectral function is defined as the trace of the imaginary part of the retarded Green function:
\begin{align}
	A(\omega,\vec{k})=\text{Tr}\left(\text{Im}\left[\mathcal{G}_R(\omega,\vec{k})\right]\right).
\end{align}
Figure ~\ref{spectralliu} shows the spectral density for this model. 
For   $b_z>M$, band crossing exists and 
the distance between the two  Weyl points becomes shorter as $M$ increases, and finally gap is open when $M>M_c$. 
Notice that  $M_c\simeq b_z$   in holography, which is   similar to  the  QFT result apart from the line broadening due to the interaction and temperature effects. See Figure \ref{spectralliu} and \ref{tempevoliu}.
However, we emphasize that the critical value of $M/b$ for the given $b$
is not exactly the same as $1$. The figure \ref{phaseD} shows the  difference.

\section{Stability of Topology}
\subsection{Topological invariants from Green function}
We study topology of holographic Weyl semi-metal (WSM) model using  the topological Hamiltonian~\cite{Wang2014,Wang2013} 
\begin{align}
	\mathcal{H}_t=-\mathcal{G}_R^{-1}(\omega=0,\vec{k}),
	\label{effH}
\end{align}
which contains  all the effects of interaction and temperature. 
We can get eigenvectors from this topological hamiltonian so that we can define Berry connection,
\begin{align}
	\mathcal{A}_{\mathbf{k}}=i\sum_j\langle n_{\mathbf{k},j}|\partial_{\mathbf{k}}| n_{\mathbf{k}j}\rangle \label{berry}
\end{align}
where $n_{\mathbf{k}}^{j}$ are eigenvectors for $\mathcal{H}_t$ in momentum space and $j$ runs over all occupied bands. The Berry phase $\gamma$ is defined by~\cite{Berry1984}
\begin{align}
	\gamma=\oint_{  \mathcal{C}}\mathcal{A}_{\mathbf{k}}\cdot d\mathbf{k}=\int_{\mathcal{S}} \mathbf{\Omega}_{\mathbf{k}}\cdot d\mathbf{S}
\end{align}
where $\mathcal{S}$ is a 2-dimensional  surface whose boundary is 
$\mathcal{C}$, a closed loop,  and $\Omega_i=\epsilon_{ijl}\left(\partial_{k_j}\mathcal{A}_{k_i}-\partial_{k_i}\mathcal{A}_{k_j}\right)$. Since the momentum space is 3-dimensional, 
we could take another surface $\mathcal{S}'$ such that its boundary is  also  $\mathcal{C}$. 
Then   the ambiguity free condition on the choice of the surface  $\mathcal{S}$ gives the condition that    
\begin{align}
	C=\frac{1}{2\pi}\oiint_{\mathcal{S}-\mathcal{S}'} \mathbf{\Omega}_{\mathbf{k}}\cdot d\mathbf{S}
	= \iiint _{B} \nabla\cdot\nabla\times \mathcal{A}_{\mathbf{k}}
\end{align}
is an interger,  a topological invariant known as Chern number. 
Here $B$ is a ball whose boundary is the closed surface $\mathcal{S}-\mathcal{S}'$. 
\begin{figure}[ht!]
\centering
    \subfigure[Spectral function ]
    {\includegraphics[width=45mm]{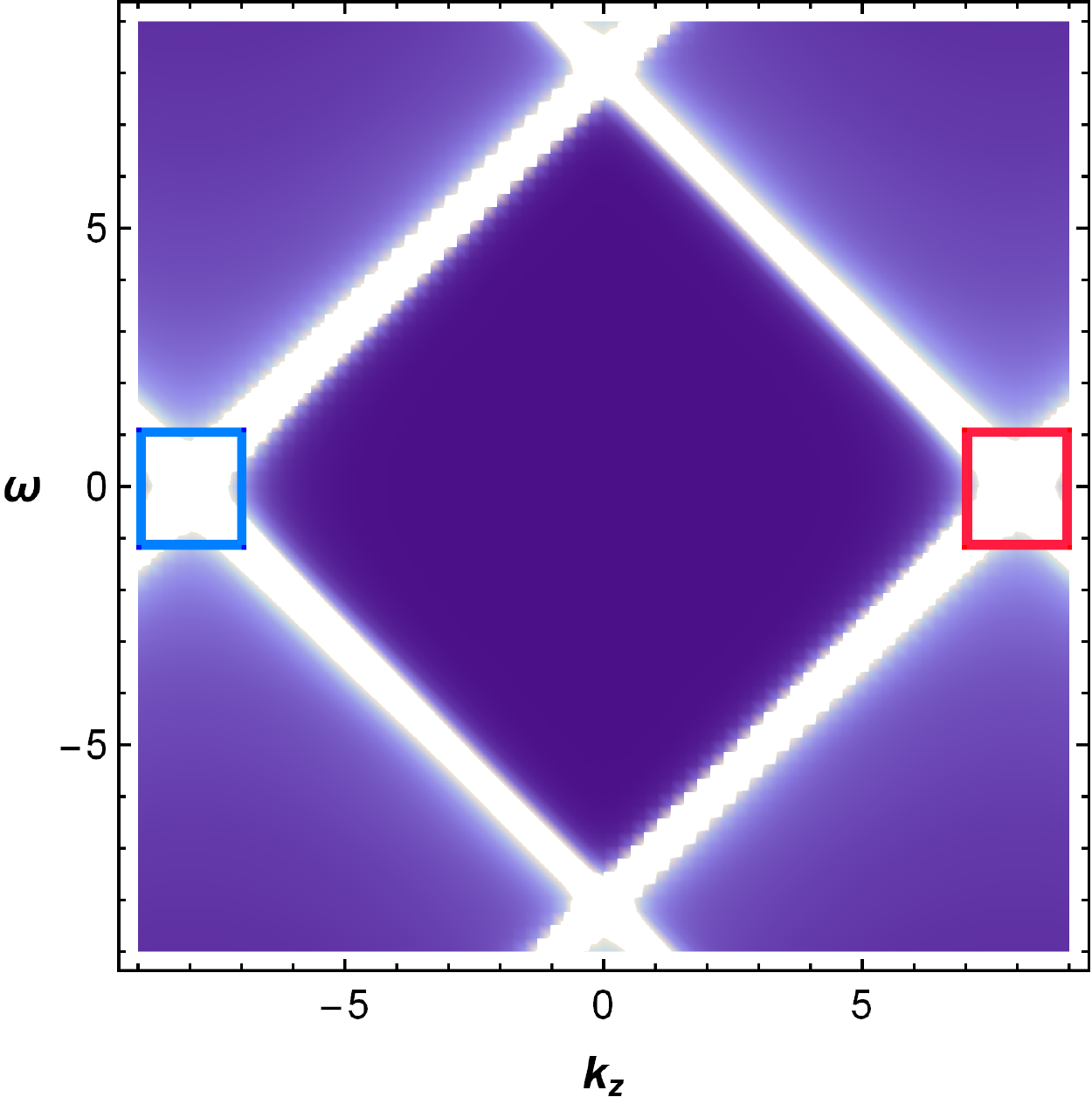}  }
    \subfigure[Berry curvatures (LHS)]
    {\includegraphics[width=53mm]{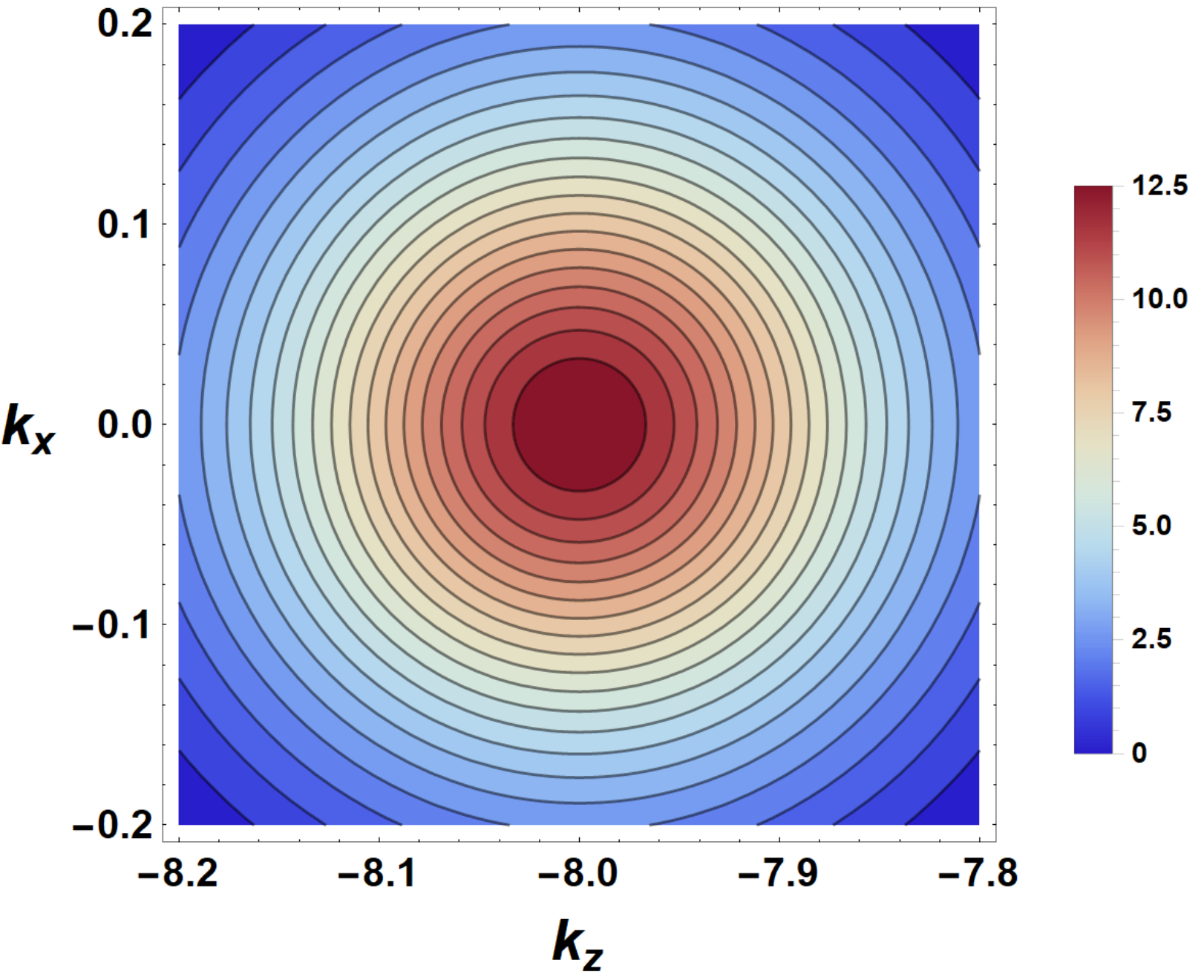}  }
      \subfigure[Berry curvatures (RHS)]
    {\includegraphics[width=53mm]{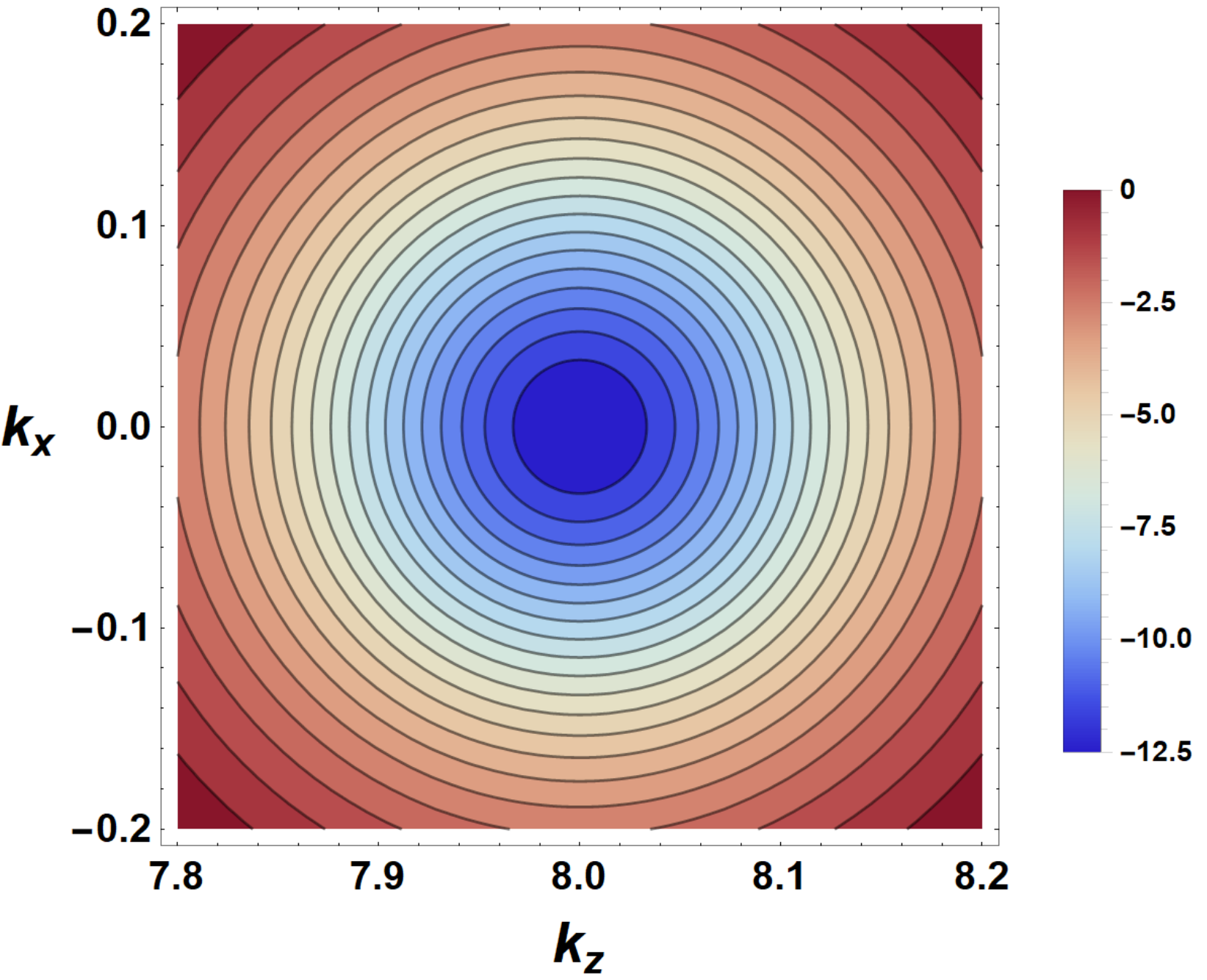}  }
              \caption{(a) Spectral function {\it without} scalar interaction. 
               We used $(b,M)=(8,0)$ and $T=2/\pi$.   $k_x=k_y=0$ fixed. 
              (b),(c): Berry curvatures near $k_{z}=-8$ and $8$ at the plane 
              ($k_{y}=-0.2$ , $\omega=0$),  where  topological numbers are  still 1 and -1 respectively.
         } \label{topocontliu}
\end{figure}
\begin{figure}[ht!]
\centering
    \subfigure[Spectral function ]
    {\includegraphics[width=45mm]{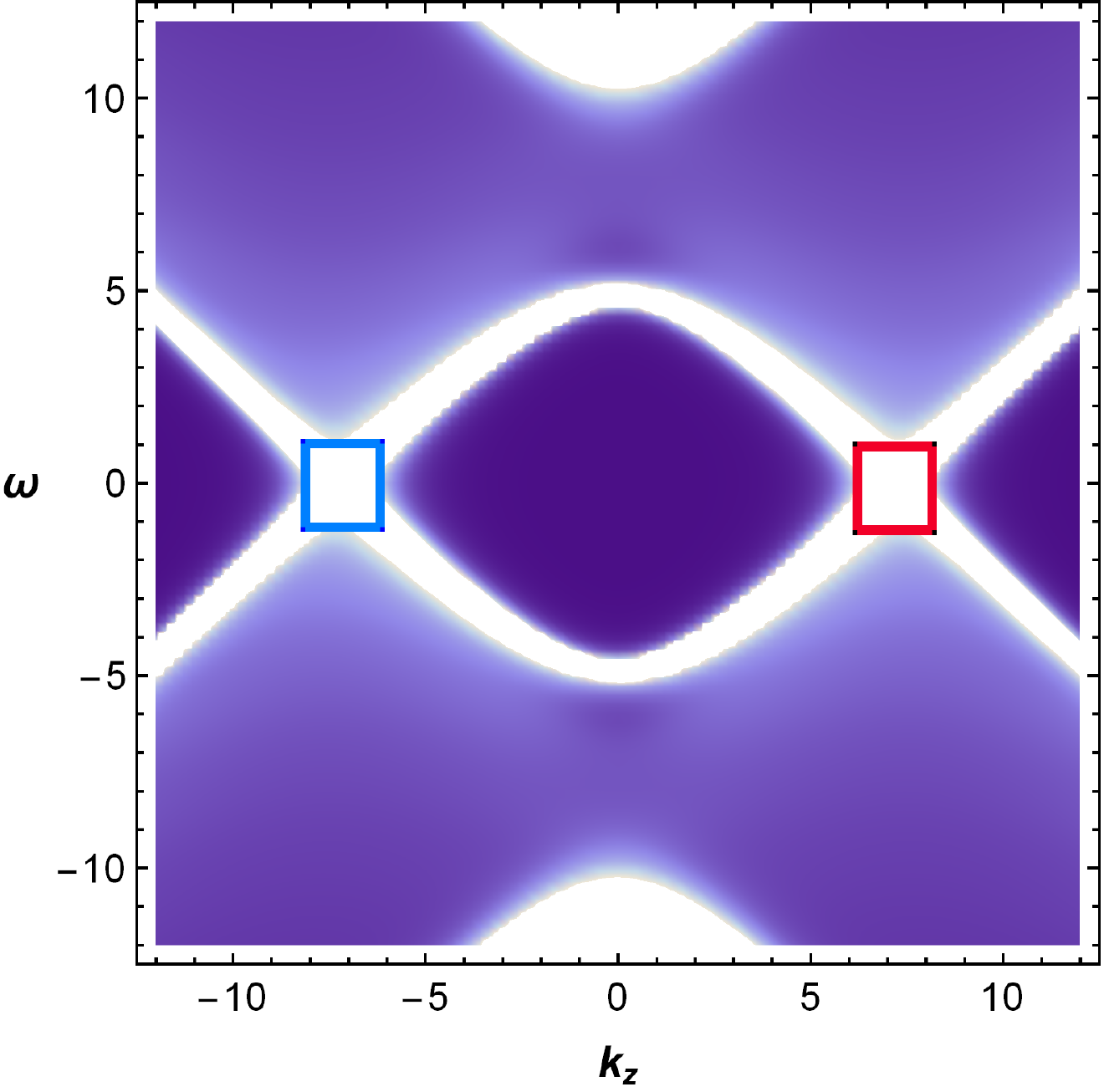}  }
    \subfigure[Berry curvatures (LHS) ]
    {\includegraphics[width=53mm]{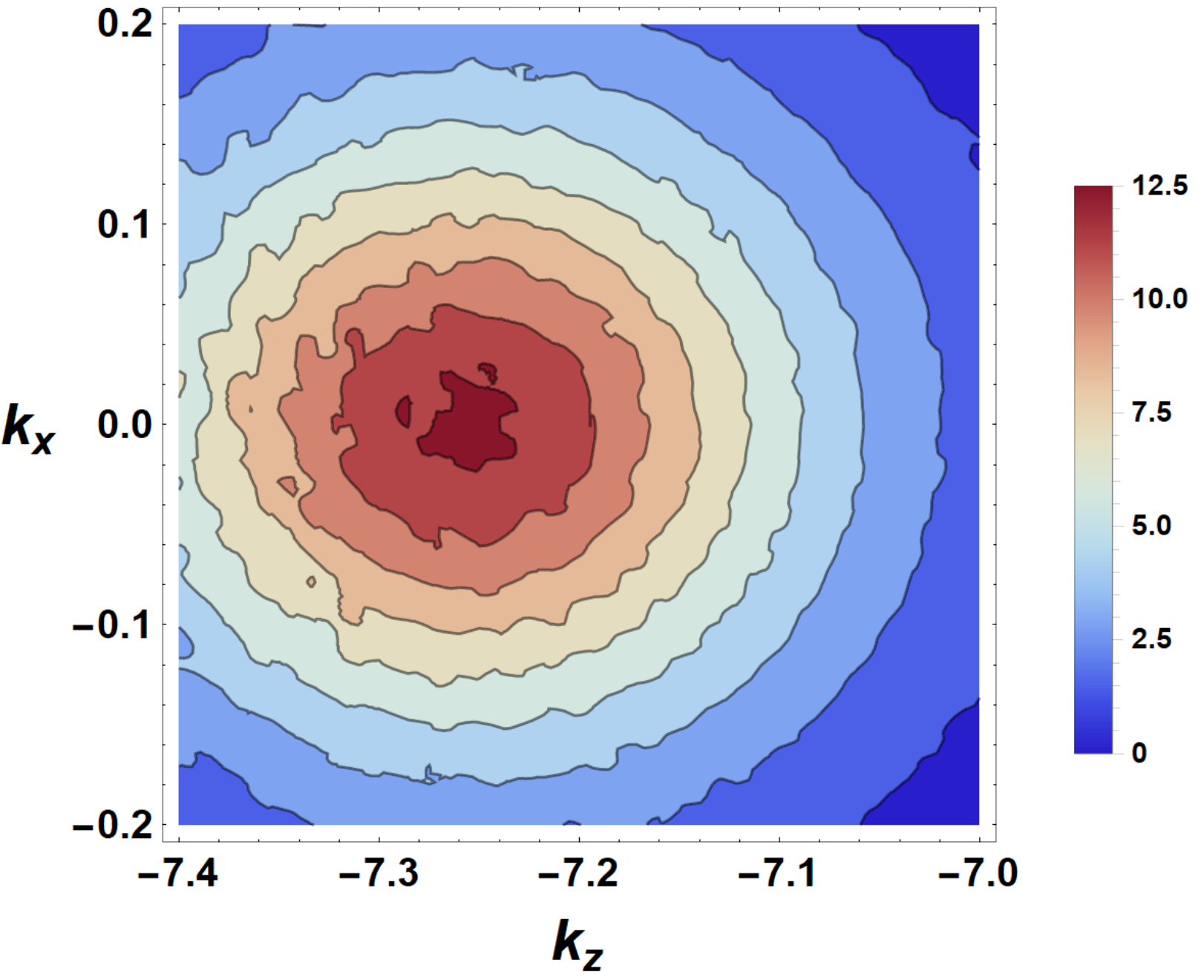}  }
      \subfigure[Berry curvatures (RHS) ]
    {\includegraphics[width=53mm]{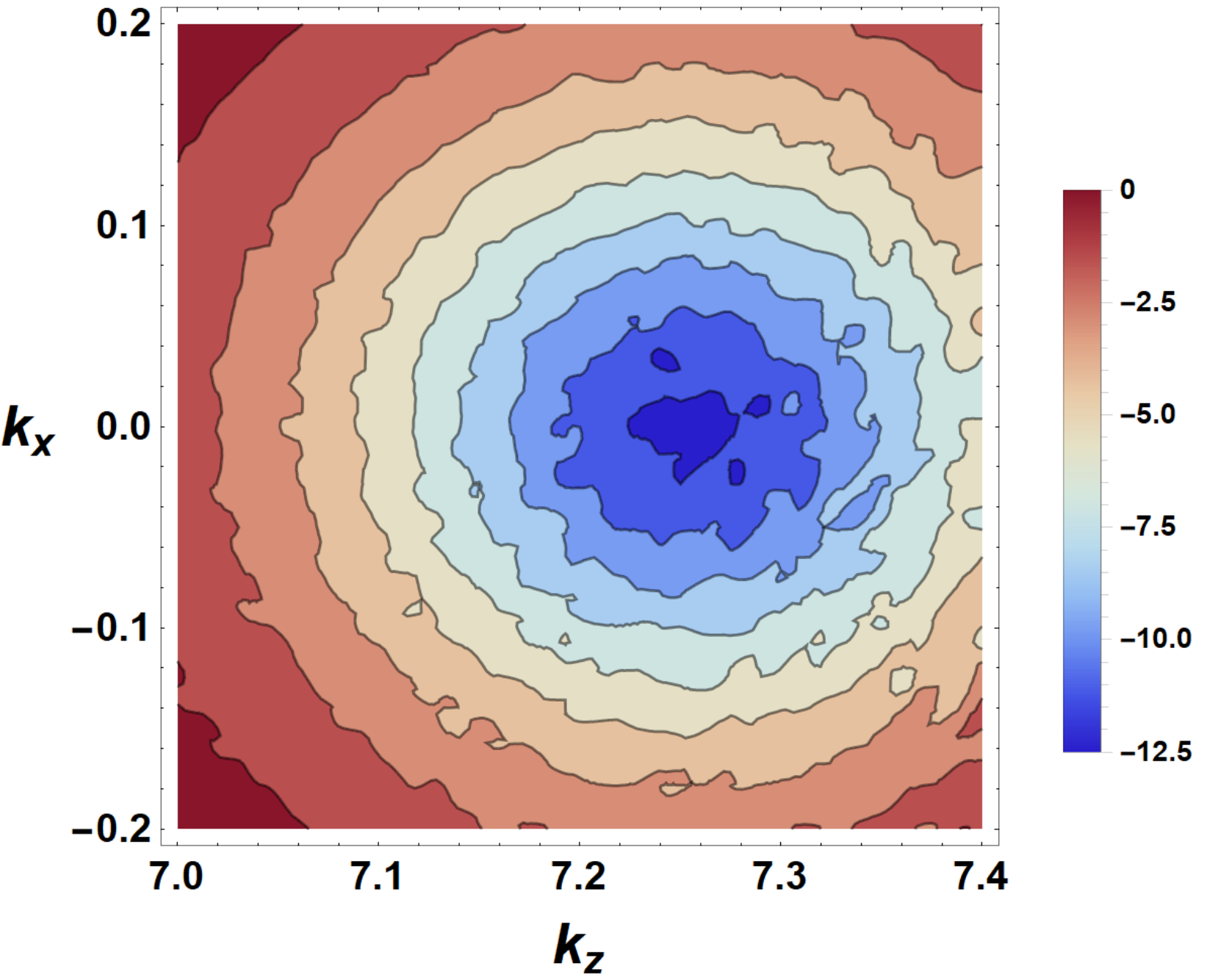}  }
              \caption{(a) Spectral function {\it with} scalar interaction.  We used
               $(b,M)=(8,5.5)$ and $T=2/\pi$.   $k_x=k_y=\omega=0$.  
               (b),(c):
               Berry curvatures near $k_{z}=-7.26$ and $7.26$ at the plane 
              ($k_{y}=-0.2$, $\omega=0$),  where  topological numbers are  still 1 and -1 respectively.
                                        } \label{topoliu}
\end{figure}

In Figure \ref{topocontliu}(a), for example, if we take   a closed surface surrounding a crossing point, then we  get  $C=1$ for the Weyl point at $k_z=-8$ and $C= -1$ for the one at $k_z=8$. Similarly, we get the  Chern numbers for $(b_{z},M)=(8,5.5)$  of Figure \ref{topoliu}. As you can see in Figure \ref{spectralliu}, the band crossing disappears  when $b_{z}<M_{c}$ which is similar to QFT case. 
In the next subsection, we will try to understand what we found here  
in more analytic terms. 

\subsection{Stability of Topology in the presence of temperature and  interaction}
Figures 7(a) and 8(a) show that  the band crossings  and  fuzziness in spectral lines simultaneously so that it is not clear whether there is a Weyl point with well defined topological number.  
Nevertheless,  figures 7(b)(c) and 8(b)(c) show that 
 there is an integer winding number for such fuzzy crossing.  To understand such numerical result found in the last subsection, 
we first notice 
%i) The band crossing is protected by the discrete symmetries, i.e, Time reversal  or Parity symmetry and this can not be changed by the presence of the interaction or temperature. The  problem is fuzziness. 
%  \\According to eq. (3.4), 
%ii) 
that the topological number of Weyl point depends only on the
local singularity structure of the Berry phase, because  in  eq. (3.4), $\nabla\cdot\nabla\times \mathcal{A}_{\mathbf{k}}$  is zero unless it is a delta function whose support is inside the ball $B$. Therefore  we only need to look at small neighborhood of the Weyl point, where only two bands are crossing. Therefore  for the purpose of the topological number,   we only   need to look at  $2\times 2$ matrix  which  describes one of the  crossing point.   This is equivalent to neglecting the  highest or lowest branches  in figure \ref{dspaxialz}(b). Any   ${2\times 2} $ matrix can be expanded in the following form 
\be
\mathcal{H}_{2\times 2}  = \vec{b}(k)\cdot\vec{\sigma}+\Sigma\mathbf{1}_2,  \label{H22}
\ee
where $\Sigma=\Sigma'+i \Sigma''$ is the self-energy.  
There are a few comments.\\
i) If $\Sigma$ were not the coefficient of the $1_{2}$ matrix, it would be a part of the momentum shift not energy shift and it would not be called as self-energy. \\
%ii)  $\Sigma$ is complex number:  $\Sigma'$  shifts the eigenvalues while $\Sigma''$ makes the line broadening and Fermi surface  fuzzy  representing the effects of interaction and temperature.   
%$\Sigma'$ can be set to be zero by shifting the chemical potential  so that the Weyl points happen always at $\omega=0$. 
ii) The       matrix becomes  non-Hermitian due to the presence of the $\Sigma''$. It is well known that  
such non-hermicity is   the result of the manybody interaction encoded in  the  1-particle effective Hamiltonian:   $\Sigma''$ is the sum of probabilities of the state of $\omega=0$  to go into all other states. \\
iii)  
 Near a crossing  point, $ \vec{b} (k) \simeq v(\vec{k}-\vec{k}_0)$ and   $k_{0}$ is real.  Equivalently,  the matrix is non-Hermitian only by the  presence of the self energy term.   

$\Sigma''$ is what makes the spectral function and the ARPES data fuzzy. Below we can easily demonstrate  the details  
why such fuzzy Fermi surface can still give well defined winding number. 
 For  simplicity, let  $v=1$ and the Weyl point be at the origin so that  $\vec{k}_{0}=0$.  The solution of eigenvalue problem 
$\mathcal{H}\ket{n_{k}}= E \ket{n_{k}} $ { are given by } 
\be 
\ket{n_{k}^{1}}=\frac{(-k_{x}+ik_{y},|\vec{k}|+k_{z})}{\sqrt{2|\vec{k}|(|\vec{k}|+k_{z})}}, \quad 
\ket{n_{k}^{2}}=\frac{(|\vec{k}|+k_{z},k_{x}+ik_{y})}{\sqrt{2|\vec{k}|(|\vec{k}|+k_{z})}} \label{eigenv}
\ee
Berry potential given by Eq.(\ref{berry}) defines  the vector potential of a magnetic  monopole sitting at $\vec{k}=0$ whose 
field strength is given by 
$
F_{12}= {k_{z}}/{2k^{3}}$
whose integral over a small sphere $S_{0}$ around the $\vec{k}=0$ is $2\pi$, so that  the    chern number $C=\int_{S_{0}} \frac{F}{2\pi}=+1$.

The key observation is that   there is no $\Sigma$ dependence in  the expression of eigenvectors in  Eq. (\ref{eigenv}). Therefore although the self energy term can make the Fermi surface fuzzy, it can not change the structure of Berry potential and hence can not change the topological structure. 
 
\vskip .1cm 

One may want to  consider  the full $4\times 4$ matrix directly instead of  looking at the crossing point,  which reduced the effective Hamiltonian to $2\times 2$ matrix.  The cost   is rather expensive: the calculation is long so that  even the result  for   the Berry potential  takes a few pages to write. 
Nevertheless we can discuss some essence of the topological structure. We describe it in the appendix C. 

%Other point to note is   the reason  for using the 
%$-\mathcal{G}_R^{-1}(0,\vec{k})$ instead of the full $\omega$-dependent  Green function:  the Weyl point is located at $\omega=0$ and the Chern number is determined by the local singularity structure of the Berry potential 
%so that knowing the Effective Hamiltonian at $\omega=0$ is good enough.  
%

\section{Topolgical dipole in a holographic theory}
In this section we consider a slightly modified model where 
some unusual but interesting phenomena happen. 
\subsection{  Spectral functions and multiple band crossing} 
We start from the topological dipole model, 
\begin{align}
	 S_1&=i\int_{\mathcal{M}}d^5x\sqrt{-g}\bar{\psi}_1\left(\Gamma^M\mathcal{D}_M-m-i A_z\Gamma^{5}\Gamma^z\right)\psi_1\nonumber\\
	 S_2&=i\int_{\mathcal{M}}d^5x\sqrt{-g}\bar{\psi}_2\left(\Gamma^M\mathcal{D}_M+m-i A_z\Gamma^{5}\Gamma^z\right)\psi_2\nonumber\\
	 S_{int}&=i\int_{\mathcal{M}}d^5x\sqrt{-g}(-\Phi\bar{\psi}_1\psi_1+\Phi\bar{\psi}_2\psi_2)\label{fermionaction}
\end{align}
where $S_{int}$ is the scalar interaction  with $\phi\sim M/r$ near boundary. We take different sign for scalar interaction $\phi$ to  make    the term invariant  under the parity transformation ($\vec{k}\to -\vec{k}$)~\cite{Semenoff2012}.\\
The equations of motion are given by
\begin{align}
	(\cancel{\mathcal{D}}-m-\Phi-i A_z\Gamma^{5}\Gamma^z)\psi_1=0\nonumber\\
	(\cancel{\mathcal{D}}+m+\Phi-i A_z\Gamma^{5}\Gamma^z)\psi_2=0
\end{align}
We can decompose the bulk fermion field into two component spionors $\psi_{I+}$ and $\psi_{I-}$, which  are eigenvectors of $\Gamma^{\underline{r}}$ with $I=1,2$ so that $\Gamma^{\underline{r}}\psi_{I\pm}=\pm\psi_{I\pm}$.
% and \begin{align}
%	\psi_I=\left(\begin{array}{c} \psi_{I+}\\ \psi_{I-}\end{array}\right)
%\end{align}
Let 
\begin{align}
	\psi_{I\pm}(r,x)=(-gg^{rr})^{-1/4}e^{ik_{\mu}x^{\mu}}\phi_{\pm}(r,k),\qquad \phi_{I\pm}=\left(\begin{array}{c}y_{I\pm}\\z_{I\pm}\end{array}\right)\label{decomp}
\end{align}
Using (\ref{decomp}), the equations of motion for bulk fermion fields $\psi_1$ are given by
\begin{align}
	\sqrt{\frac{g_{ii}}{g_{rr}}}y_{1+}'(r) -(m+\Phi)\sqrt{g_{ii}}y_{1+}(r)+(u_+-k_z)y_{1-}(r)-(k_x-ik_y)z_{1-}(r)=&0\nonumber\\
	\sqrt{\frac{g_{ii}}{g_{rr}}}z_{1+}'(r) -(m+\Phi)\sqrt{g_{ii}}z_{1+}(r)-(k_x+ik_y)y_{1-}(r)+(u_-+k_z)z_{1-}(r)=&0\nonumber\\
	\sqrt{\frac{g_{ii}}{g_{rr}}}y_{1-}'(r) +(m+\Phi)\sqrt{g_{ii}}y_{1-}(r)-(u_++k_z)y_{1+}(r)-(k_x-ik_y)z_{1+}(r)=&0\nonumber\\
	\sqrt{\frac{g_{ii}}{g_{rr}}}z_{1-}'(r) +(m+\Phi)\sqrt{g_{ii}}z_{1-}(r)-(k_x+ik_y)y_{1+}(r)-(u_--k_z)z_{1+}(r)=&0 \label{eom}
\end{align}
where $u_{\pm}=\omega\pm A_z$. One can get the equations of motion  for $\psi_2$ by $m,\Phi\to -m,-\Phi$, which changes the chirality.
At the boundary region ($r\rightarrow \infty$), the geometry becomes asymptotically $AdS_5$, so that Eqs. (\ref{eom}) have asymptotic solution as
\begin{align}
	y_{1+}(r)&=A_{11} r^m+B_{11} r^{-m-1}, \qquad y_{1-}(r)=C_{21} r^{m-1}+D_{11} r^{-m}\nonumber\\
	z_{1+}(r)&=A_{12} r^m+B_{12} r^{-m-1}, \qquad z_{1-}(r)=C_{22} r^{m-1}+D_{12} r^{-m}  \label{asymptotics}\\
\quad 	y_{2-}(r)&=A_{21} r^m+B_{21} r^{-m-1}, \qquad y_{2+}(r)=C_{21} r^{m-1}+D_{21} r^{-m}\nonumber\\
	z_{2-}(r)&=A_{22} r^m+B_{22} r^{-m-1}, \qquad z_{2+}(r)=C_{22} r^{m-1}+D_{22} r^{-m}\label{asymptotics2}
\end{align}
Here, we have two independent sets of equations. Each set needs four initial conditions, but as in WSM, we can fix half of them by choosing infalling condition at the horizon. Hence, it is required that we choose two independent initial conditions so that we can obtain two corresponding sets of source and expectation values to compute Greens function for each $\psi_I$. By denoting each initial conditions as (1), (2) respectively, we can construct the source and expectation matrices as $2\times 2$ matrices. 
\begin{align}
	\mathbf{A}_I=\left(\begin{array}{cc}
		A_{I1}^{(1)} &A_{I1}^{(2)}\\
		A_{I2}^{(1)} &A_{I2}^{(2)}
	\end{array}\right),\qquad 
	\mathbf{D}_I=\left(\begin{array}{cc}
		D_{I1}^{(1)} &D_{I1}^{(2)}\\
		D_{I2}^{(1)} &D_{I2}^{(2)}
	\end{array}\right)
\end{align}
The retarded Green function is defined by $\mathcal{G}_I^R=i\gamma^t\mathbf{D}_I\mathbf{A}_I^{-1}=-\mathbf{D}_I\mathbf{A}_I^{-1}$. However, since we know that each set of Greens function is independent, so we can construct Greens function matrices as $4\times 4$ block diagonalized matrices which is given by 
\begin{align}
	\mathbf{G}^R=\left(\begin{array}{cc}
		\mathcal{G}_1^R &0\nonumber\\
		0 &-\mathcal{G}_2^R
	\end{array}\right)
\end{align}
where the $-$ sign in front of $\mathcal{G}_2^R$ represents the alternative quantization~\cite{Laia2011}.
\begin{figure}[ht!]
\centering
    \subfigure[$(b,M)=(8,5.5)$ ]
    {\includegraphics[width=35mm]{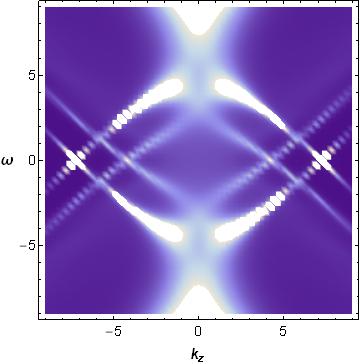}  }
    \subfigure[$(b,M)=(8,7.5)$ ]
    {\includegraphics[width=35mm]{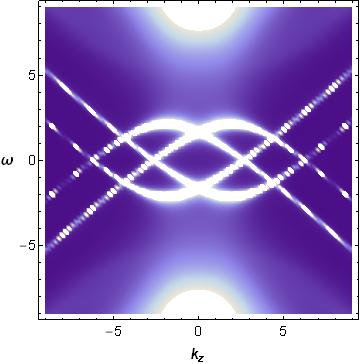}  }
     \subfigure[$(b,M)=(8,11)$ ]
    {\includegraphics[width=35mm]{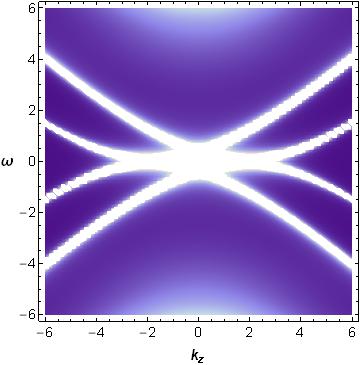}  }
     \subfigure[$(b,M)=(8,12.7)$ ]
    {\includegraphics[width=35mm]{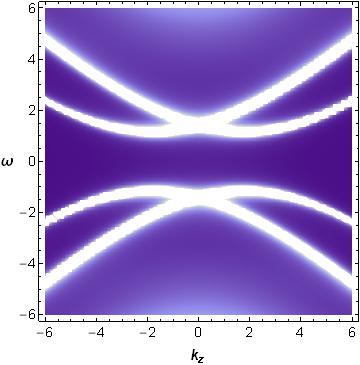}  }
              \caption{(a)-(d): Band structure in  $(k_z,\omega)$  with $k_x=k_y=0$ of model in Eq. (4.1). We used $T=2/\pi$.
              Notice that at the critical point (c), it is not reduced to a Dirac 
point but to a higher Lifshitz point.         } \label{comparison}
\end{figure}

%\newpage
Comparing these  spectral functions with holographic WSM case, 
one can see from  Figure \ref{comparison} that the outermost part of the spectrum evolves  similarly to that of WSM case. As $M$ increases, the separation between outermost band-crossing points decreases and, after $M>M_c=4.3$, a gap opens and its size  gets larger. However, there are crucial difference is that here we have multiple band-crossings.   
Each band forms cone-like structure.  

\begin{figure}[ht!]
\centering
    \subfigure[$T=2/\pi$ ]
    {\includegraphics[width=45mm]{b80M55di_t2pi.jpg}  }
   \subfigure[$T=10/\pi$ ]
    {\includegraphics[width=45mm]{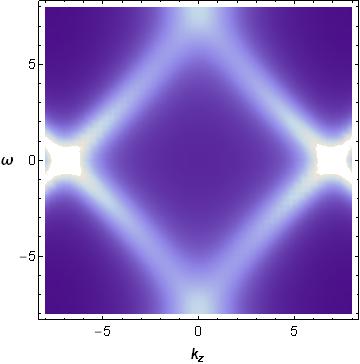}  }
     \subfigure[$T=20/\pi$ ]
    {\includegraphics[width=45mm]{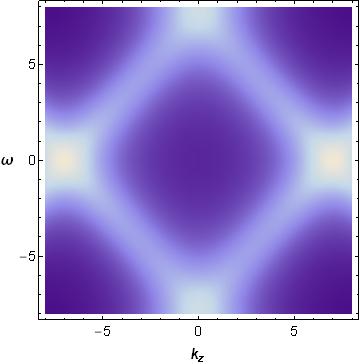}  }
              \caption{(a)-(c):Temerpature effect on the spectral function when $(b,M)=(8,5.5)$ on $(k_z,\omega)$-space with $k_x=k_y=0$. 
         } \label{tempevo}
\end{figure}

As we increase the temperature, the spectrum goes fuzzy and the distance between adjacent spectra also increases with the position of the outermost part of spectrum fixed, which implies the number of crossing points near $\omega=0$ decreases. See Figure \ref{tempevo}.
\subsection{Topological dipoles }
\begin{figure}[ht!]
\centering
    \subfigure[Spectral function ]
    {\includegraphics[width=40mm]{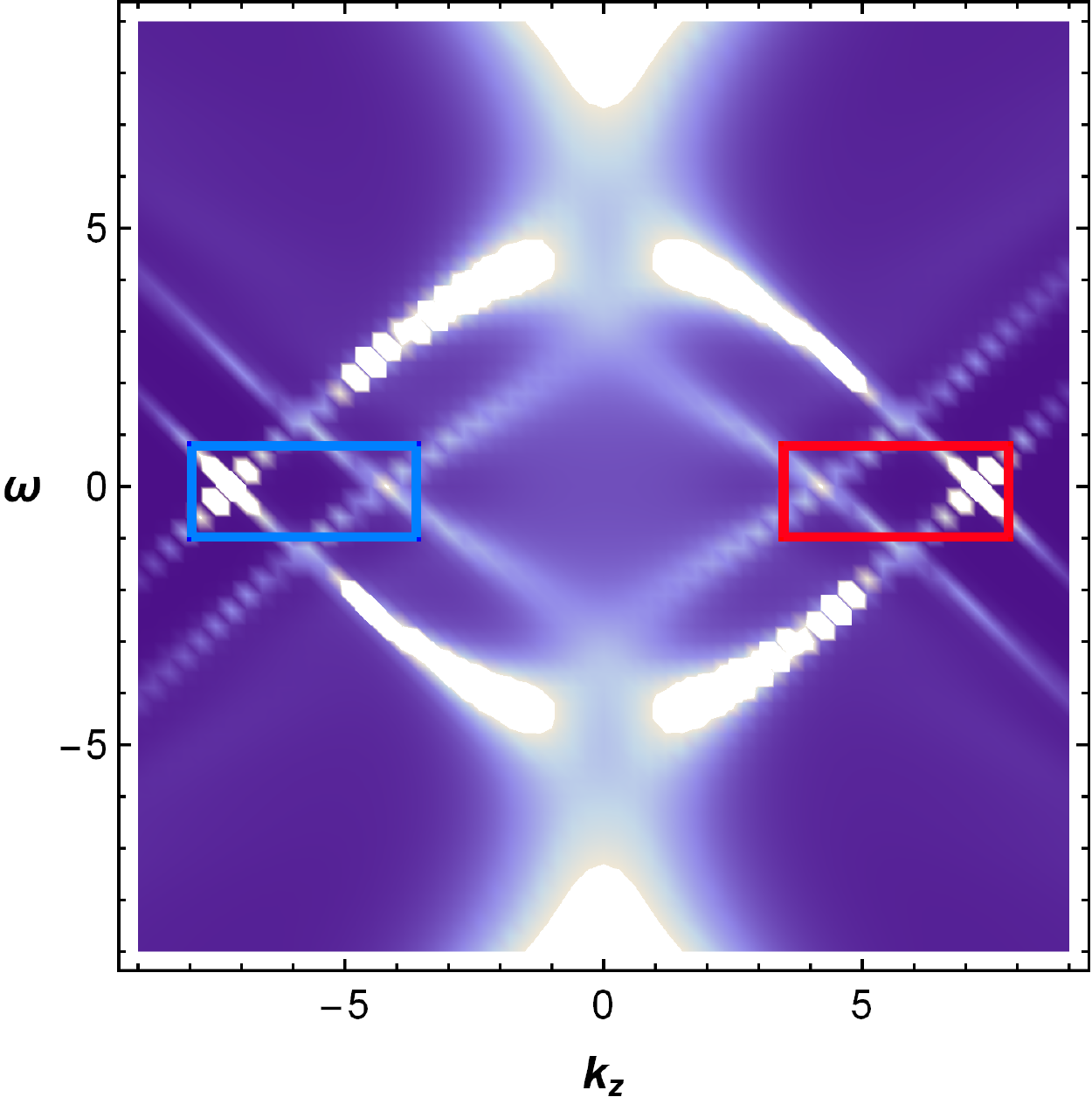}  }
    \subfigure[Berry curvatures ($k_{z}<0$) ]
    {\includegraphics[width=55mm]{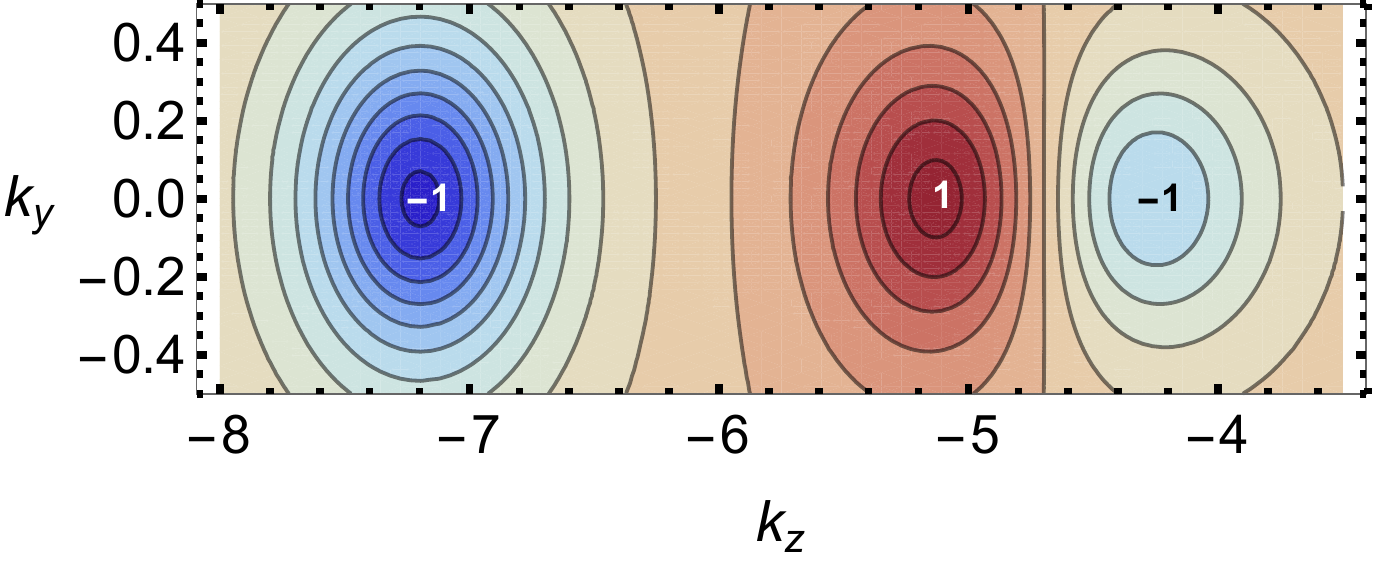}  }
    \subfigure[Berry curvatures ($k_{z}>0$)  ]
    {\includegraphics[width=55mm]{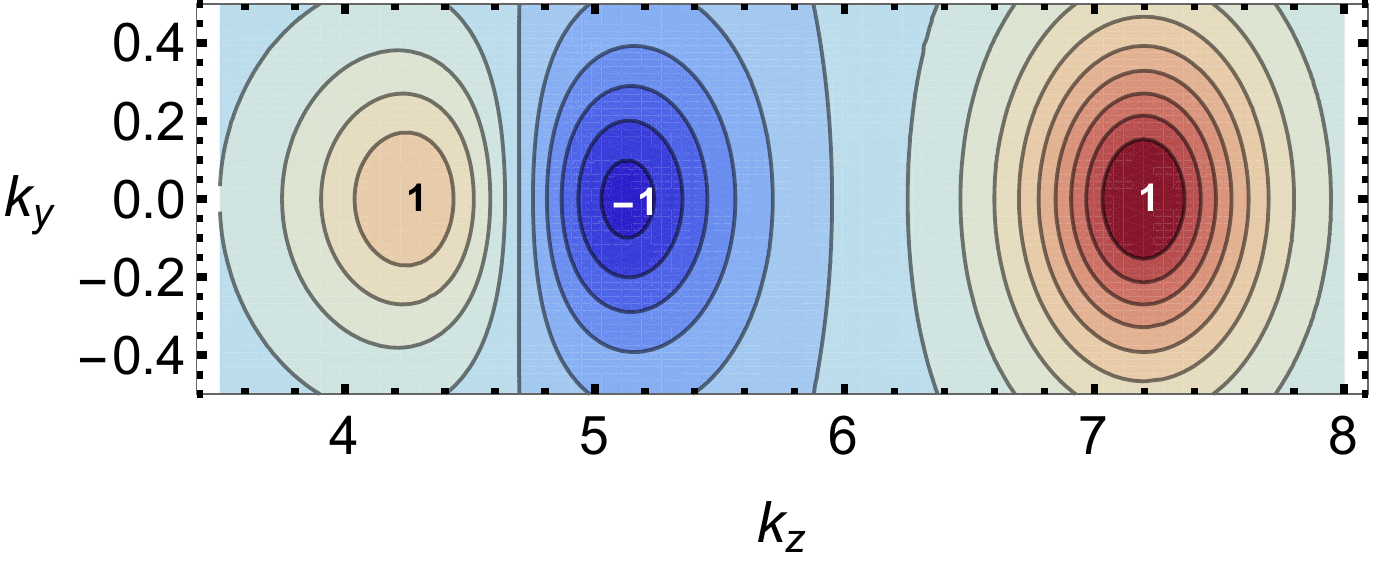}  }
              \caption{(a) Spectral function with $k_x=0,k_y=0$. Colored boxes are the regions where we  draw Berry curvatures  at Fermi level in (b) and (c) with 
               $(b,M)=(8,5.5)$ and $T=2/\pi$. Each numbers in (b) and (c) represents winding  number computed from Berry curvatures. As we can see from Fig \ref{tempevo}, this dipole structure can disappear when temperature goes up. 
         } \label{topocont}
\end{figure}

We can also calculate  the topological invariants of crossing points for this model. In this case, we have multiple crossing points  unlike  WSM, therefore we calculate Berry phase at each crossing points. For $k<0$, net topological invariants near the band crossing point are -1 while those for $k>0$ are 1. However, except for the outermost  band crossing points, each poles of  Berry curvature 
comes with its conjugate pair with opposite sign  to  make dipoles. 
See Figure \ref{topocont}. Notice that as we increase the temperature, the inner band-crossing points disappear (Figure \ref{tempevo}). Since they depend on temperature sharply, we may consider that they are not stable. This is not surprising since two topological charges are so closely separated in momentum space, it is expected to be unstable for relatively small perturbations.  
From the Schr\"odinger potential picture, each band is induced by a confining well structure and disappearance  in temperature means that the potential well is eaten by the black hole horizon as temperature increases.   
 Notice that the leftmost  crossing point  has topological charge of $(-1,2)$ with small separation: it is a  combination of the Weyl point having topological charge +1 and small separated dipole charges (-1,1). \\
 To explain this better, we draw the contribution of $\psi_{1}$ and $\psi_{2}$ separately. For $\mathcal{G}_1^R$, there are 4 band crossing points. 
 The rightmost one is the Weyl point with topological charge -1 and other three crossing points  are dipoles with small separated  -1 and +1 from left to right. 
Similarly for $\mathcal{G}_2^R$, there are 4 band crossing points and 
 the leftmost one is the Weyl point with topological charge +1 and other three crossing points  are dipoles with small separated  -1 and +1 from left to right. 
Now combining these two,  the position of dipole's +1 charge happens to coincide with that  of the Weyl point with charge +1. \\
Similar story goes on for the rightmost band crossing with opposite monople charge and the same dipole charge.
 
Why dipole should come in this model?  We can understand it as follows.  Two fermions $\psi_{1} , \psi_{2}$ are not mixing directly. 
So if one fermion has total Weyl charge +1,  the other one has -1.  Now if we draw spectral function for $\psi_{1}$, the shape has multiple crossings.  Each crossing point has well defined topological charge.  Now, the right-most crossing has -1 and unpaired, therefore  all   other 3 crossing points  should come as a pair  with topological charge  (+1,-1)  not to change the total charge $-1$.  
Similarly  for $\psi_{2}$, the left-most crossing has +1 and unpaired, therefore every other ones  should come as pairs. 
This is the reason why dipole should appear. 
Notice that if we sum two spectral function, then the total charge is summed to be 0.     See figure \ref{topocont}.

\section{Discussion}
In this paper, we discussed the stability of the topological invariant of the interacting Weyl semi-metal
at finite temperature and finite fermion mass.
We utilize the holographic setup to  calculate the Green's functions and use the result to construct the effective Hamiltonian, which allows us to calculate the winding number of the Weyl points.  
We found that the topological winding number is stable even in the case where spectral function is fuzzy. The winding number turns out to be integer as far as there is band crossing at the Fermi level. 
Here we summarize the arguments why that is so: 
the topological number's integrand is $ddA_{Berry}$ which is zero or delta function by Bianchi identity  so that we only need to look at neighborhood of the Weyl point, where only two bands are crossing. Therefore it is enough to consider  2x2 matrix only. 
The interaction can change the spectral function to make it fuzzy by creating $\Sigma''$ but the latter  is a coefficient of $\bf 1_{2}$ so that  it can not modify the Eigenvectors which are the building blocks of the Berry potential determining the winding number. 
The formula of winding  number is nothing but the reading machine of coefficient of that singularity, which can not be changed by smooth deformation of the theory  by  the interaction or temperature.

One subtle point at this moment is  whether the interaction can develop the imaginary part of $b$ in Eq.(3.5). 
We numerically checked that the expression is still $b\sim (k-k_{0})$ near the crossing point.  In fact, if there is a  tiny  imaginary part in $k_{0}$,  one  can show that topological number is zero. If a weak interaction or small temperature can induce imaginary part in $k$, it is equivalent to saying that 
 topology is unstable for small deformation, which does not make sense.   
  So we believe that the reality of $k_{0}$ is protected by a discrete  symmetry.  It is equivalent to say that effective Hamiltonian can be non-Hermitian only by the presence of imaginary self-energy term which is diagonal. In fact,  there is no reason why interaction can generate arbitrary non-hermitian structure in the effective Hamiltonian. We  want to comeback to the analysis of various discrete symmetry   in the future. 
   
We also defined a model where Weyl points are separated only by a small distance. We call it Topological Dipoles and study its topological invariant. 

It would be very interesting to generalize this work to the case with more general type of interaction and also to Dirac materials in 2+1 dimension  having other type singularities. 
For example, for the line node  cases,  multiple  band crossings can define a new type of topological invariant\cite{Ahn:2018qgz}. 
New type of topological matter called ``Fragile Topology'' \cite{Po:2017uom,PhysRevLett.120.266401,Hwang:2019nuf} is also interesting possibility. 
We hope we comeback to these issues in future works.  

\vskip 1cm
\appendix
%\begin{center}
\section*{\bf \Large Appendix}
%\end{center}
\section{Effective Schr\"odinger potential}
\begin{figure}[ht!]
\centering
       \subfigure[$V_{\text{eff}}$ for different interaction]
    {\includegraphics[width=65mm]{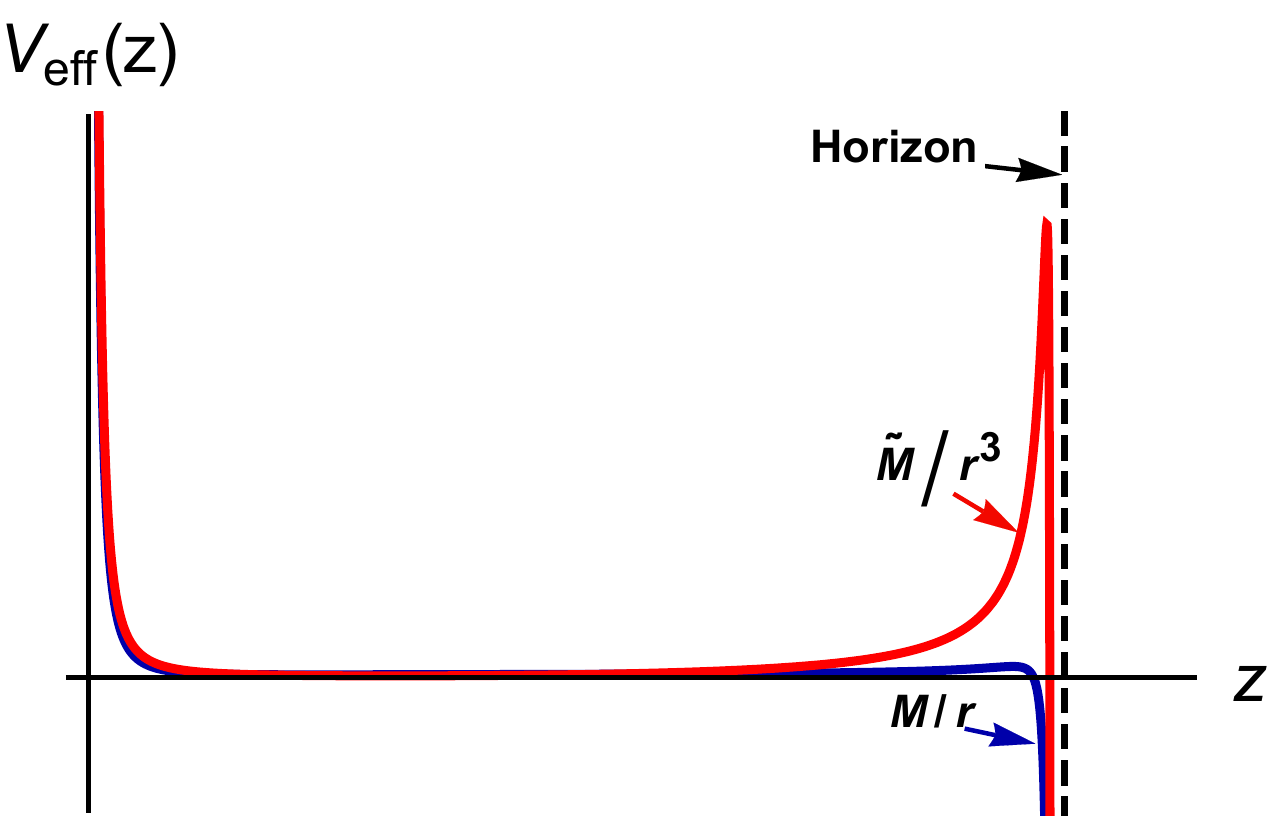}  }
     \subfigure[Temperature evolution of $V_{\text{eff}}$]
    {\includegraphics[width=65mm]{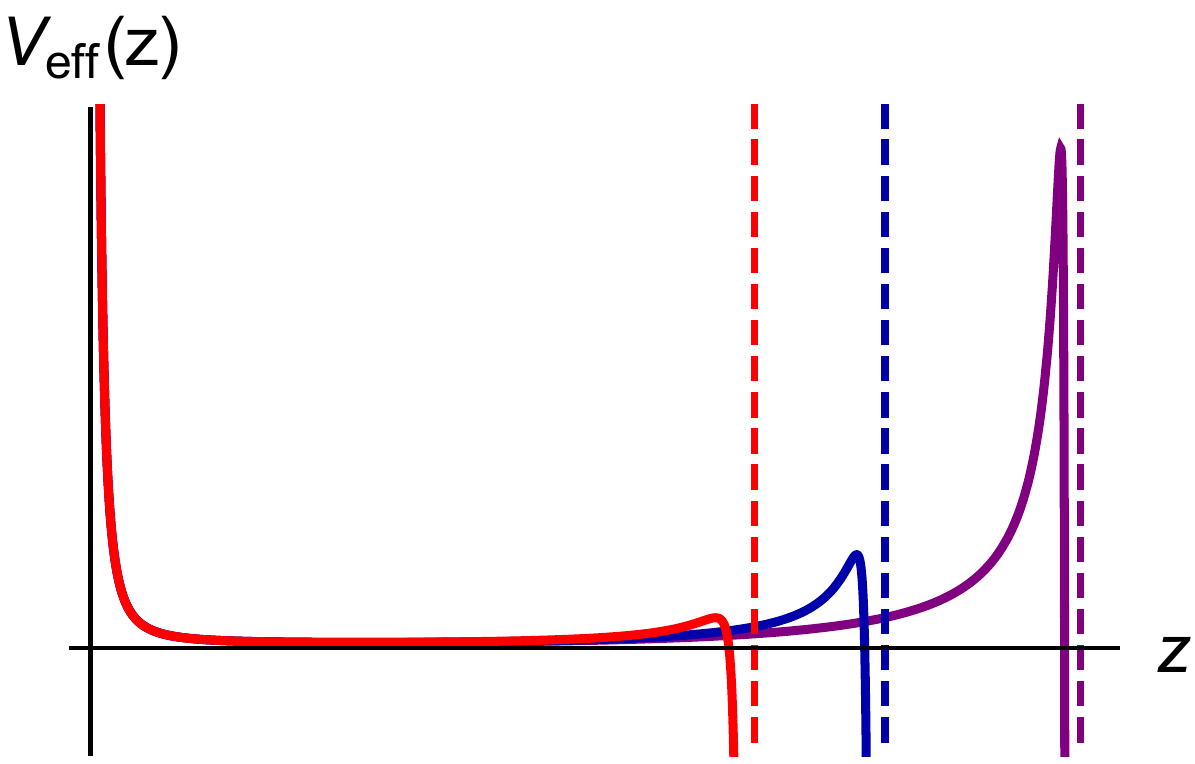}  }
              \caption{Schematic Schroedinger  potential : (a) Comparison between $M/r$ and $\tilde{M}/r^3$. The coordinate used is $ z=1/r$. (b) Temperature evolution of the effective potential in the bulk. Dashed lines are the  event horizons at  each temperature which moves in as T increases, and the potential changes from purple to red accordingly. 
         } \label{potential}
\end{figure}

We can define the effective bulk potential to analyze the role of scalar interaction for Dirac fermion in the bulk.
\begin{align}
S_D=\int d^5x \sqrt{-g}\bar{\psi}(\Gamma^M\mathcal{D}_M-m-\phi)\psi+S_{bd}
\end{align}
Here, we use the pure AdS background which is given by
\begin{align}
	ds^2=-r^2dt^2+\frac{dr^2}{r^2}+r^2d\vec{x}^2
\end{align}
The Dirac equation in Fourier space is given by
\begin{align}
	r\Gamma^r\partial_r\psi+\frac{i}{r}\Gamma\cdot k\psi+2\Gamma^r\psi-(m+\phi)\psi=0\label{diffscalar}
\end{align}
We can decompose the bulk fermion field into two component spionors $\psi_+$ and $\psi_-$, which correspond to eigenvectors of $\Gamma^{\underline{r}}$. That is,  $\Gamma^{\underline{r}}\psi_{\pm}=\pm\psi_{\pm}$ and 
$ 	\psi^{T}=\left(  \psi_+, \psi_- \right) $.\\
Then, equation (\ref{diffscalar}) becomes coupled equations for $\psi_{\pm}$
\begin{align}
	\psi_+=-\frac{i\gamma\cdot k}{k^2}A(-m,-\phi)\psi_-,\qquad \psi_-=\frac{i\gamma\cdot k}{k^2}A(m,\phi)\psi_+
\end{align}
where $k^2=\vec{k}^2-\omega^2$ and
\begin{align}
	A(m,\phi)=r\left(r\partial_r+2-m-\phi\right)\label{diffscalar2}
\end{align}
from which we obtain ~\cite{Iqbal2009}
\begin{align}
	k^2\psi_+=A(-m,-\phi)A(m,\phi)\psi_+
\end{align}
Changing the coordinate $r=1/z$, we substitute $\psi_+=r^2\psi$ to the equation \ref{diffscalar2}. Then, we can get the Schr\"odinger form of the Dirac equation:
\begin{align}
	-\psi''&+V_{eff}(z)\psi_n=E\psi\nonumber\\
	V(z)&=\phi(z)+\frac{m(m-1)}{z^2}
\end{align}
We can extend the analysis to the finite temperature case, which has the Schwarzschild-$AdS_5$ background. We will not show the details of calculation for this since it is  very complicated. We just show the  effective bulk potential schematically for finite temperature case in Figure \ref{potential}. As the temperature increases, the radius of black hole horizon increases so that the effective potential cannot form steep wall near the horizon. Hence, there cannot be a bound state, which gives a band structure.       

But, we cannot apply this analysis to our model because we have the momentum transfer in spectral densities and it might not be possible to construct the Schrodinger form of equations of motion. Even if it is possible to construct, we need to encode the momentum dependence on the effective potential, which should be done for each fixed momentum. 

For the simple scalar interaction, there's no momentum dependence,   it is enough to calculate the case of $k=0$.  Since $r$ corresponds to energy scale and $k=0$ means small energy scale so that the contribution of $V_k(r)$ at $r\sim0$ is significant. On the other hands, for our model, which does have momentum dependence, the k dependence of the effective potential is significant. Hence the behavior of $V_k(r)$ at the finite $r$ is important. 
%It will be discussed in detail later.

\section{Holographic Greens function}
We start from the simple probe fermion model, 
\begin{align}
	S_{bulk}=\int_{\mathcal{M}}d^5x\sqrt{-g}i\bar{\psi}\left(\Gamma^M\mathcal{D}_M-m)\right)\psi\label{fermionaction2}
\end{align}
However, the action (\ref{fermionaction2}) is not sufficient and the boundary term is required to guarantee that the variational principle is well-defined. We will discuss later. 
The equations of motion are given by
\begin{align}
	(\cancel{\mathcal{D}}-m)\psi=0
\end{align}
Taking decomposition of the bulk fermion field into two component spinors  $\psi_+$ and $\psi_-$ as previous section, we now expand the bulk-spinors in Fourier-space as following: 
\begin{align}
	\psi_{\pm}(r,x)=(-gg^{rr})^{-1/4}e^{ik_{\mu}x^{\mu}}\phi_{\pm}(r,k),\qquad \label{decomp2}.
\end{align}
Setting  $\phi_{\pm}^{T}=\left( y_{\pm}, z_{\pm} \right)$,  
the eqs. of motion for bulk fermions are given by
\begin{align}
	\sqrt{\frac{g_{ii}}{g_{rr}}}y_+'(r) -m\sqrt{g_{ii}}y_+(r)+(\omega-k_z)y_-(r)-(k_x-ik_y)z_-(r)=&0\nonumber\\
	\sqrt{\frac{g_{ii}}{g_{rr}}}z_+'(r) -m\sqrt{g_{ii}}z_+(r)-(k_x+ik_y)y_-(r)+(\omega+k_z)z_-(r)=&0 \\
 \hbox{and } \qquad
	\sqrt{\frac{g_{ii}}{g_{rr}}}y_-'(r) +m\sqrt{g_{ii}}y_-(r)-(\omega+k_z)y_+(r)-(k_x-ik_y)z_+(r)=&0\nonumber\\
	\sqrt{\frac{g_{ii}}{g_{rr}}}z_-'(r) +m\sqrt{g_{ii}}z_-(r)-(k_x+ik_y)y_+(r)-(\omega-k_z)z_+(r)=&0 \label{eom2}, 
\end{align}
using (\ref{decomp2}). 
Near the boundary ($r\rightarrow \infty$), the geometry becomes asymptotically $AdS_5$ spacetime, so that the equations of motion (\ref{eom2}) have asymptotic behaviors as
\begin{align}
	y_+(r)&=A_1 r^m+B_1 r^{-m-1}, \qquad y_-(r)=C_1 r^{m-1}+D_1 r^{-m}\nonumber\\
	z_+(r)&=A_2 r^m+B_2 r^{-m-1}, \qquad z_-(r)=C_2 r^{m-1}+D_2 r^{-m}\label{asymptotics2}
\end{align}
Back to the boundary term, we take the variation to the bulk action (\ref{fermionaction2}),
\begin{align}
	\delta S_{bulk}=\int_{\partial\mathcal{M}}d^4x\sqrt{-h}\frac{i}{2}(\delta\bar{\psi}_+\psi_-+\bar{\psi}_-\delta\psi_+-\delta\bar{\psi}_-\psi_+-\bar{\psi}_+\delta\psi_-)+\text{bulk part}
\end{align}
where $h=gg^{rr}$ and bulk part vanishes when the Dirac equations holds~\cite{Laia2011}.\\
Since the Dirac equation is the  first order differential equation, we cannot fix both $\psi_+$ and $\psi_-$ on the boundary simultaneously. Therefore we need the additional boundary to give well-difined variational principle:
\begin{align}
	S_{bdy}=\pm\frac{i}{2}\int_{\partial\mathcal{M}}d^4x\sqrt{-h}\bar{\psi}\psi=\pm\frac{i}{2}\int_{\partial\mathcal{M}}d^4x\sqrt{-h}(\bar{\psi}_-\psi_++\bar{\psi}_+\psi_-)\label{bdy}
\end{align}
The sign can be determined such that we take positive(negative) sign when we fix the value of $\psi_+$($\psi_-$) at the boundary, where $\delta S_{bdy}$ cancels all the terms including $\delta\psi_-$ ($\delta\psi_+$) in $\delta S_D$. And this defines the standard(alternative) quantization. By using (\ref{asymptotics2}), the boundary action in (\ref{bdy}) becomes
\begin{align}
	S_{bdy}\sim y_-z_-+y_+z_+=(A_1D_1+A_2D_2)+\Sigma_{\pm}E_{\pm}r^{\pm 2m-1}+E_2r^{-2},
\end{align}
It seems that $S_{bdy}$ blows up at the boundary when $m>1/2$, but, it can be cancelled by introducing proper counter terms~\cite{Ammon2010}, which do not have finite terms at the boundary. As we mentioned above, if we choose the standard quantization, we should fix $\psi_+$ at the boundary so that $A_i$ are the sources and $D_i$ are the expectation values. From now on, we will hold to this quantization rule. Therefore, if variables with $-$ index and those with $+$ in equations are not mixed such as $k_x=k_y=0$,   the retarded Green's function is given by
\begin{align}
	\mathcal{G}=\text{diag}\left(-\frac{D_1}{A_1},-\frac{D_2}{A_2}\right)\equiv\text{diag}(G_+,G_-)
\end{align}
In this paper, however, we should consider all $(\omega,\vec{k})$ space so that the variables in equations cannot be decoupled. Hence, we need to define the Green function in another way. We have 4 variables and each needs 1 initial condition, hence 4 initial conditions are required. By choosing infalling functions at the horizon, we can relates $y_+$ to $y_-$ and $z_+$ to $z_-$, which means there are only two dimensional space of initial condition. These are the pair of coefficients of infalling wave functions of $y_+,z_+$ : Denote them $(y_+^0,z_+^0)$. Two independent basis vector of this space can be chosen as $(1,1)$ and $(1,-1)$. Let's call them $\vec{e_1}$ and $\vec{e_2}$ respectively. For each $i$, $\vec{e_i}$ determines their counterpart at the horizon ($i.e$ $y_-^0,z_-^0$)  and one can integrate the equations of motion from the horizon to the boundary. Then, for each initial conditions $\vec{e_i}$, $i=1,2$, we get near-boundary solution
\begin{align}
	y_+^{(i)}&=A_1^{(i)}r^m+B_1^{(i)}r^{-m-1},\qquad y_-^{(i)}=C_1^{(i)}r^{m-1}+D_1^{(i)}r^{-m}\nonumber\\
	z_+^{(i)}&=A_2^{(i)}r^m+B_2^{(i)}r^{-m-1},\qquad z_-^{(i)}=C_2^{(i)}r^{m-1}+D_2^{(i)}r^{-m}\label{asym2}
\end{align}
Since the choice of coefficients is arbitary, the general boundary solutions should be a linear combnination of the such solution with some coefficients. For example, $y_+=c_1 y_+^{(1)}+c_2 y_+^{(2)}$. For the general boundary solutions, its coefficients $X_a$, where $X=A,B,C,D$ are given by $X_a=\sum_iX_a^{(i)}c_i$. The matrix $\mathbf{X}$ is defined by the components $X_a^{(i)}$ where $a,i=1,2$ are the row index and columb index respectively. We can solve for $c_i$: by definition, $\vec{c}=\mathbf{A}^{-1}\vec{A}$ for $X=A$. Green function can be derived from the relation between $\vec{D}$ and $\vec{A}$:
\begin{align}
	\vec{D}\equiv\mathcal{S}\vec{A}=\mathbf{D}\mathbf{A}^{-1}\vec{A}\quad \Rightarrow \quad\mathcal{G}_R\equiv i\gamma^t\mathcal{S}=-\mathbf{D}\mathbf{A}^{-1}
\end{align}
% \newpage
\section{Topology in 4 by 4 Hamiltonian model for WSM}
First  let's  assume following form of the effective Hamiltonian: 
\begin{align}
	\mathcal{H}  &=-\mathcal{G}_R^{-1}(0,\vec{k})
	=
	\left(\begin{array}{cc}\vec{b}_{1}(k)\cdot\vec{\sigma}+\Sigma\mathbf{1}_2   & m\mathbf{1}_2\\ m\mathbf{1}_2 & \vec{b}_{2}(k)\cdot\vec{\sigma}+\Sigma\mathbf{1}_2   \end{array}\right)
	  \label{topH}
\end{align}
where $ \vec{b}_{1}(k) \simeq v(\vec{k}-\vec{k}_0)$   near    the right Weyl point and 
 $\vec{b}_{2}(k)\simeq -v(\vec{k}+\vec{k}_0)$ near the left Weyl point. 
 In fact,   if we start from (\ref{Action}),   the effective hamiltonian   Eq.(\ref{effH}) numerically calculated always takes   above  form. 
 There are a few steps to prove that the presence of topological number. 
\begin{enumerate}
\item For small $m$ and near $k_{0}$, taking the expansion in $m$ gives us  
  $\vec{B}=\curl \vec{A}(k)=\frac1{2}\frac{\vec{p}}{|p|^{3}}+ {\cal O}{(m^{2})}$ with $p=k-k_{0}$. % where $k_e=\sqrt{k_0^2-m^2}$. 
\item $\nabla\cdot \vec{B} =0$  off the Weyl point   by the Bianchi identity,. 
\item From 1 and 2,  $\nabla\cdot \vec{B} = 2\pi\delta(\vec{k}-\vec{k_{0}})$.   Therefore  $C=\frac1{2\pi}\int_{S_{0}} B\cdot dS $ is non-zero integer 
or zero depending on whether $S_{0}$ contains Weyl point  $k_{0}$ or not. 
\item The topological number $C$ is independent of $m$,  therefore we can set $m=0$ for the purpose of calculating the Chern number.  
 For  $m=0$,  we only need to handle $2\times 2$ matrix, which was already done above. 
\end{enumerate}

One can prove   that   topological structure is intact as far as  $k_{0}$ is real. 
The fact that $k_{0}$ does not get imaginary part  is due to the  discrete symmetry P (parity) and  T(Time reversal).  Even in the case where the Fermi sea  disappears  due to the    interaction, such crossing point   is located  exactly at the $\omega=0$ by T symmetry. When $k_{0}$ gets imaginary  numbers, the monopole singularity is smoothed out  in the real domain and the chern number becomes 0. 
 This is expected because   the topology of a manifold and the singularity  of the harmonic function defined on it is   equivalent and because 
  the singularity of the  monopole field $\sim 1/k^{2}$ is resolved.       

Note  that  the  figure \ref{spectralliu} is for finite temperature. 
The position of the Weyl point is not changed compared with the Minkowski space, the non-interacting case, but there is finite line broadening.  Figures \ref{topocontliu} and \ref{topoliu} are the spectral functions 
at different temperature and for different fermion mass $M$, where one can see that the  topological number is  the same integer value in spite of  very different broadening widths. 
Increasing temperature makes the spectral lines even  fuzzier leaving the topological invariant still fixed. 

 Therefore the topological structure is very stable  under the variation of temperature and  interactions in  holographic theory.
The argument here can be generalized to other class of topological matter although 
we  focus here on Weyl semi-metal Hamiltonian. 

In appexdix C, we consider more general cases where 
$m \mathbf{1}_2$ is replaced by $m\mathbf{1}_2+\vec{q}\cdot\vec{\sigma}$ and classify the cases where Weyl points   exist. 

%In the main text, we have shown that we can construct Hamiltonian for WSM as follows: 
%\begin{align}
%	\mathcal{H}_0=
%	\left(\begin{array}{cc}(\vec{p}-\vec{p_0})\cdot\vec{\sigma}  & m\mathbf{1}_2 \\ m\mathbf{1}_2&-(\vec{p}+\vec{p_0})\cdot\vec{\sigma}   \end{array}\right)
%\end{align}
%where $\vec{p}=(k_x,k_y,k_z)$ and $\vec{p_0}=(0,0,k_0) $. We have non-trivial topological numbers in each Weyl point at $\pm\vec{p_0}$ if $m$ is  small enough ($m<k_0$).  

Now, what if the off-diagonal interaction  is added to the
$\mathcal{H}_0$
so that the  effective Hamiltonian becomes  
\begin{align}
	\mathcal{H}=
	\left(\begin{array}{cc}(\vec{p}-\vec{p_0})\cdot\vec{\sigma},  & m\mathbf{1}_2  +\vec{q}\cdot \vec{\sigma} \\ m\mathbf{1}_2+\vec{q}\cdot \vec{\sigma} , &-(\vec{p}+\vec{p_0})\cdot\vec{\sigma}   \end{array}\right)
\end{align}
where $\vec{q}=(q_1,q_2,q_3)$. 
For  Weyl points,  Eigenvalues  should vanish  only at two distinct points with $k_x=k_y=0$. It is difficult to get explicit form of Eigenvalues of $\mathcal{H}$ in general. 
Therefore we  classify the cases and study one by one.
\subsection{$m=0$}
\begin{itemize}
\item $q_3=0$\\
When $q_3=0$ and $q_1,q_2\neq 0$, the dispersion relation is given by 
\begin{align}
	(\omega-k_z)^2=k_0^2+q_1^2+q_2^2
\end{align}
which implies that two Weyl points always exist at $(k_{x},k_{y},k_z)=(0,0,\pm\sqrt{k_0^2+q_1^2+q_2^2})$.
See figure \ref{q1q2curve}(a). 
\begin{figure}[ht!]
\centering
    \subfigure[$(k_0,q_1,q_2)=(2,1,2)$ ]
    {\includegraphics[width=45mm]{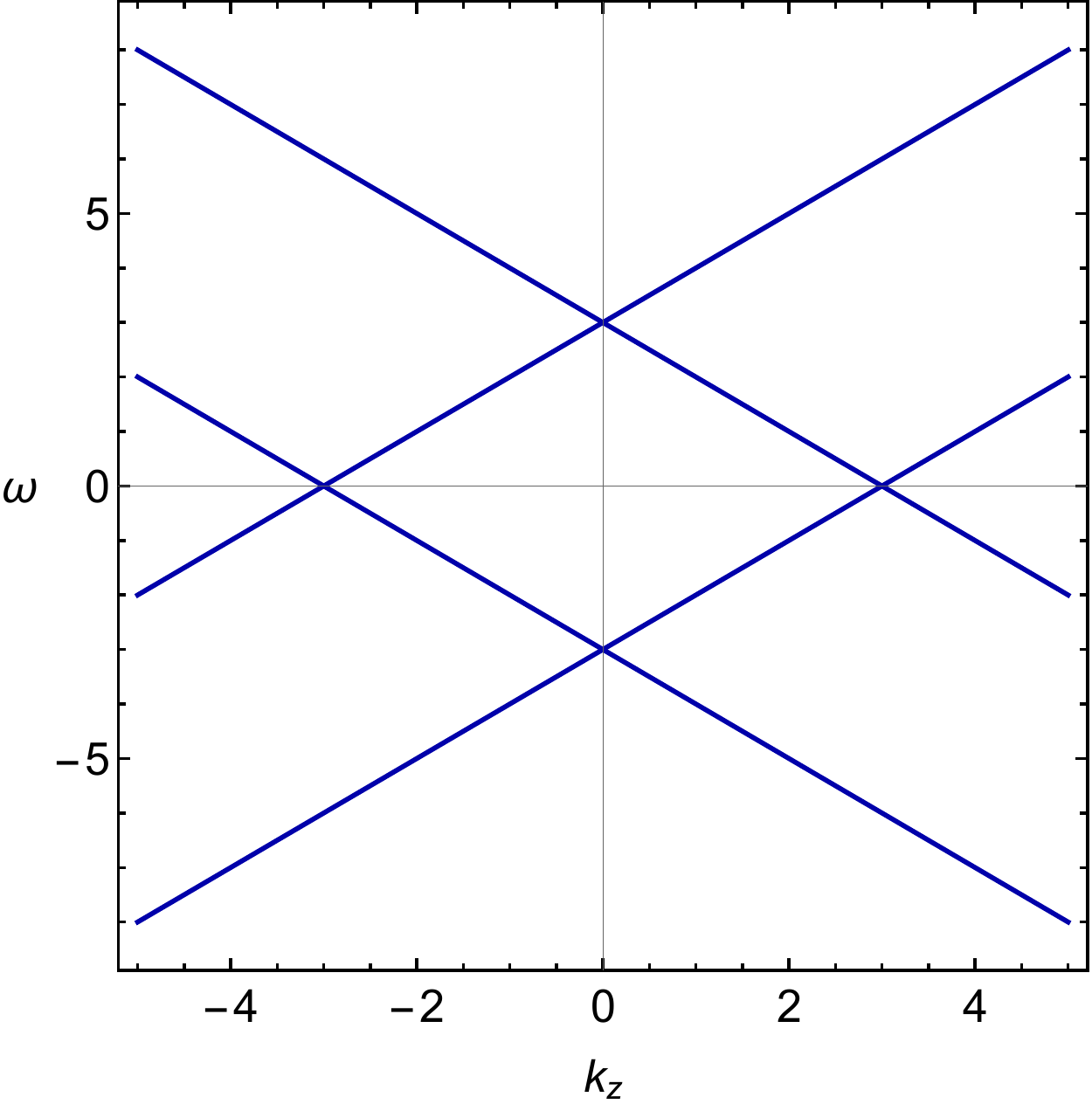}  }
    \hskip 1cm
        \subfigure[$(k_0,q_1,q_3)=(2,1,2)$ ]
    {\includegraphics[width=45mm]{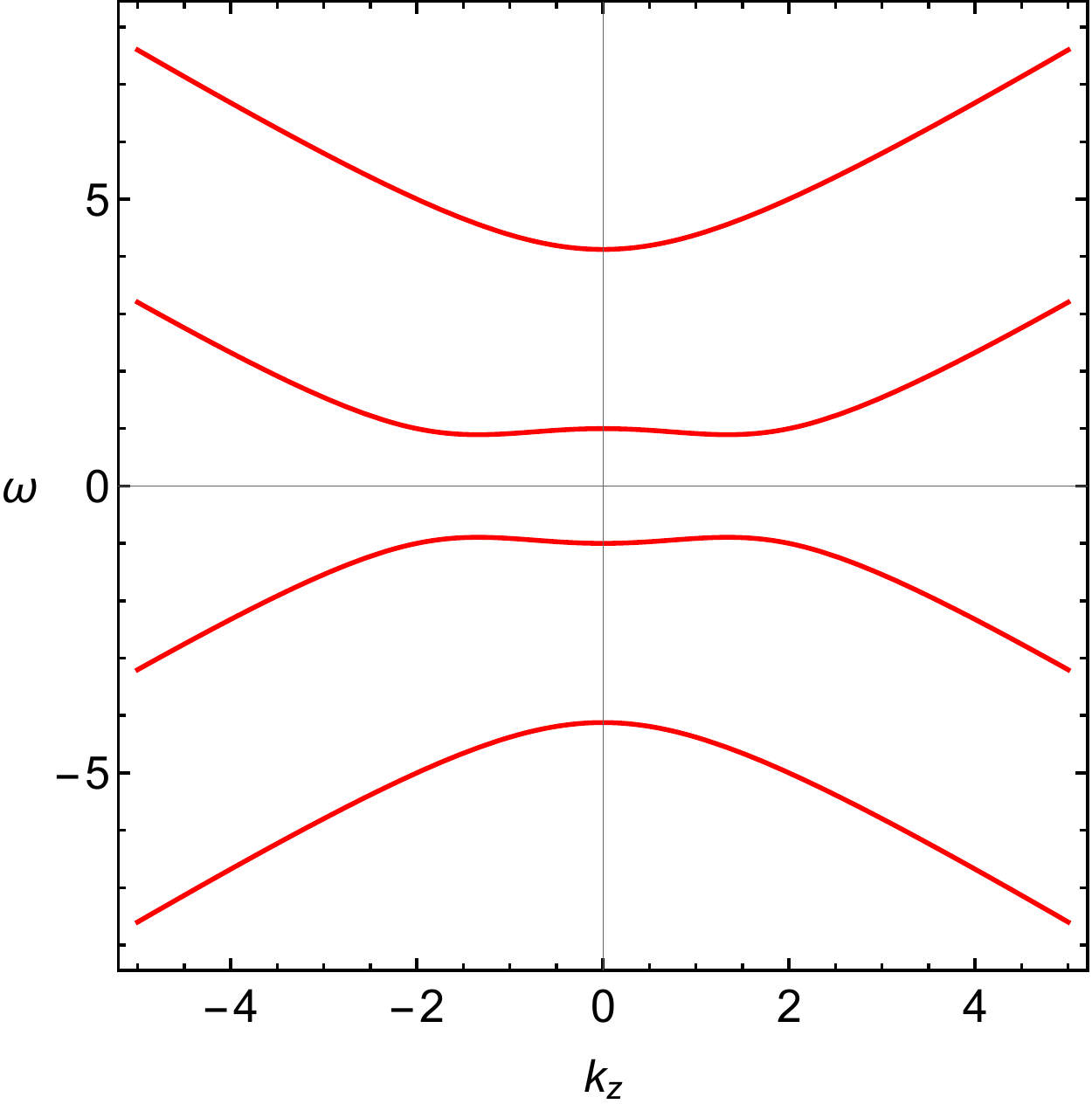}  }
                 \caption{Dispersion curve for $m=0$ 
                 (a) $q_1,q_2\neq 0$, (b) $q_1,q_3\neq 0$ 
         } \label{q1q2curve}
\end{figure}

\item   $q_3\neq 0$ and $q_1q_2=0$ \\
When $q_2=0$, the dispersion relation  is given by 
\begin{align} \omega^2=k_z^2+k_0^2+q_1^2+q_3^2\pm2\sqrt{k_z^2(k_0^2+q_1^2)+k_0^2q_3^2}
\end{align}
In this case, there is no Weyl point and this hamiltonian is gapped. See figure \ref{q1q2curve}(b). 
This result is symmetric  for $q_1$ and $q_{2}$.
One should notice that only when $q_{1}=q_{2}=0$ one can have Weyl point in the presence of $q_{3}$. 

\item $q_3\neq 0$ and $q_1=q_2=0$,  the dispersion relation becomes 
\begin{align} \omega =\pm\Big(k_0 \pm\sqrt{k_z^2 + q_3^2}\Big). 
\end{align}
In this case,   the system has two Weyl points if $k_{0}>q_{3}$, or  gapful if $k_{0}<q_{3}$. 
\end{itemize}

\subsection{$m\neq0$}
\begin{figure}[ht!]
\centering
    \subfigure[$(m,k_0,q_1,q_2)=(1,2,1,2)$ ]
    {\includegraphics[width=45mm]{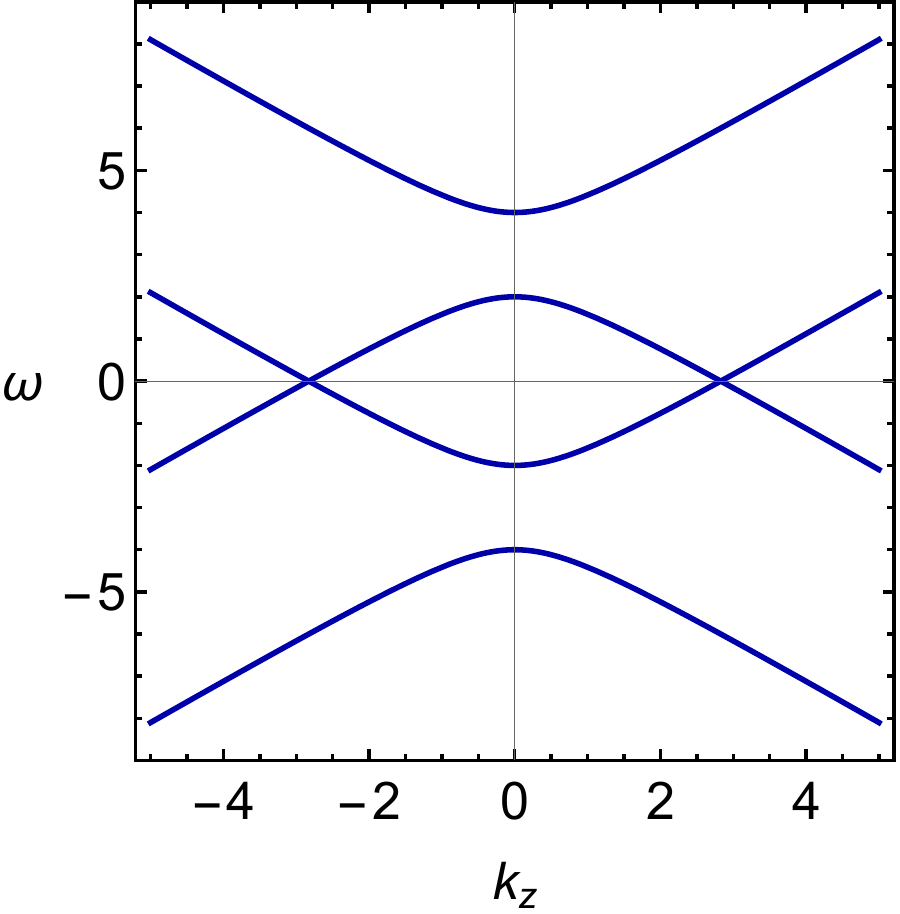}  }
 \hskip 1cm 
        \subfigure[$(m,k_0,q_3)=(2,5,3)$ ]
    {\includegraphics[width=45mm]{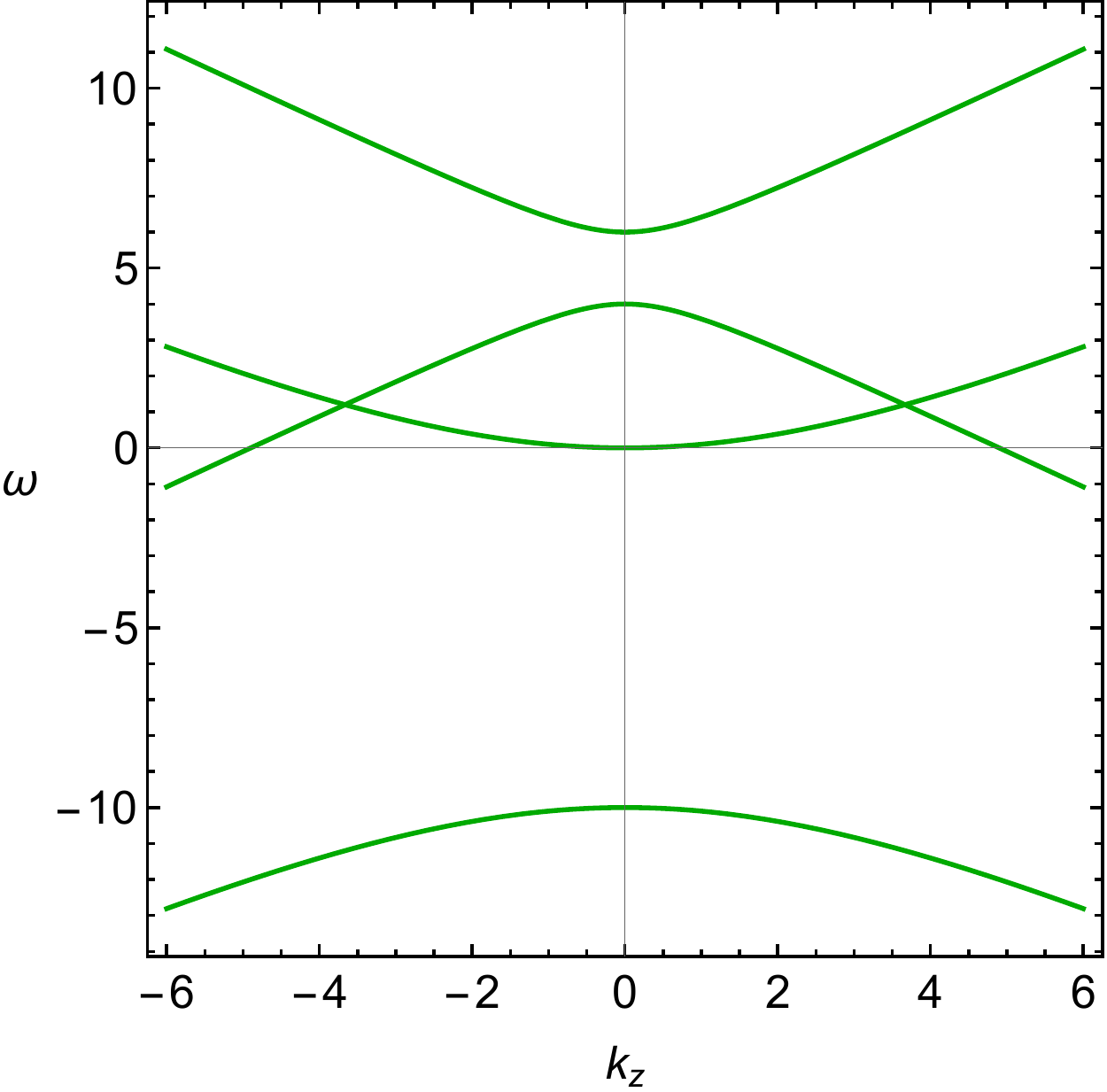}  }
                 \caption{Dispersion curve for $m>0$. 
                (a)  $q_1,q_2\neq 0$ and (b) $q_1,q_2=0$
         } \label{mq1q2curve}
\end{figure}

When both $m$ and $q_{3}$ are non-zero we can not get analytic expression for the dispersion curve unless $q_{1}=q_{2}=0$. 
\begin{itemize}
\item {$q_3=0$}
\end{itemize}
When $q_3=0$, the dispersion curve is given by
\begin{align}
\omega=\pm\left(\sqrt{k_z^2+m^2}\pm\sqrt{k_0^2+q_1^2+q_2^2}\right)
\end{align}
which implies that Weyl points exist at $k_z^2=k_0^2-m^2+q_1^2+q_2^2 $ if   $k_0^2+q_1^2+q_2^2 >m^2$ holds and gapful otherwise. See figure \ref{mq1q2curve}(a)

\begin{itemize}
\item {$q_{3}\neq 0$ and $q_1=q_2=0$}
\end{itemize}
When $q_1=q_2=0$, the dispersion curve is given by
\begin{align}
	(\omega-k_0)^2=k_z^2+(m\pm q_3)^2
\end{align}
In this case, we have 4 roots for $\omega=0$ and as you can see from dispersion curve,  the Weyl points are lifted due to the shift of $\omega$ by $k_{0}$. 
See figure \ref{mq1q2curve}(b). 
If both  $m$ and $q_{3}$ are non-zero, we could get  analytic expression for the dispersion curve only when $q_{1}=q_{2}=0$. Therefore we leave it future work for such case. 

%by $\omega\rightarrow \omega-k_0-\abs{k_0-\frac{m q_3}{k_0}}$.

\acknowledgments
% \begin{acknowledgements}
 This  work is supported by Mid-career Researcher Program through the National Research Foundation of Korea grant No. NRF-2016R1A2B3007687.  
%\end{acknowledgements}
\bibliographystyle{JHEP}
 \bibliography{Refs_dipole.bib} 
 \end{document}